\documentclass[12pt]{article}
\usepackage{amssymb,amsmath,amsthm,amsfonts,amscd}
\usepackage{graphicx}
\usepackage{subfigure}
\usepackage{color}
\definecolor{springgreen}{rgb}{0, 0.4, 1}
\usepackage[colorlinks=true, linkbordercolor={1 0 1},citecolor=springgreen]{hyperref}
\newcommand\be{\begin{equation}}
\newcommand\ee{\end{equation}}
\newcommand\bea{\begin{eqnarray}}
\newcommand\eea{\end{eqnarray}}

\newcommand{\fatsigma}{{\bf \sigma \kern -0.54em \sigma}}
\newcommand{\tpchi}{{\bf \chi \kern -0.35em \chi}}
\newcommand{\llambda}{{\bf \lambda \kern -0.45em \lambda}}

\bibliography{plain}
\title{\bf Geometric phase for two-mode entangled squeezed-coherent states}
\author{ S. Mohammadi Almas $^{a}$
 \thanks{sanazmohammadi@uma.ac.ir} ,G. Najarbashi $^{a}$ \thanks{Najarbashi@uma.ac.ir},  A. Tavana $^{a}$ \thanks{Tavana@uma.ac.ir}\\
\\ \\
$^a${\small Department of Physics, Faculty of Science, University of Mohaghegh Ardabili. P. O. Box 179, Ardabil, Iran.}\\
{\small}} \pagebreak

\begin{document}
\maketitle
\newpage 
\begin{abstract}
\par
In this paper, we study the geometric phase (GP) of two-mode entangled squeezed-coherent states (ESCSs), undergoing a unitary cyclic evolution. It is revealed that by increasing the squeezing parameter of the first or the second mode of a balanced ESCS, the GP compresses in an elliptical manner along the axis of the coherence parameter of the corresponding mode. While in the case of unbalanced ESCS, the GP compresses in a hyperbolic manner by increasing the squeezing parameters of either mode. By generalizing to higher constituting-state dimensions, it is found that the GPs of both balanced and unbalanced ESCSs, increase for a specific value of the coherence parameter. Based on these findings, using the interferometry approach, we suggest a theoretical scheme for the physical generation of the balanced ESCS.
\par
 {\bf PACs Index: }
\end{abstract}
\section{Introduction}
\par
The real-Abelian geometric phase (GP), was first discovered by Berry \cite{berry1984quantal} for pure quantum states undergoing cyclic adiabatic evolutions. It has already been generalized in many different aspects, e.g., non-Abelian GP\cite{wilczek1984appearance}, non-adiabatic \cite{aharonov1987phase} and noncyclic \cite{samuel1988general, mukunda1993quantum} evolutions, mixed states \cite{uhlmann1986parallel, sjoqvist2000geometric} and open systems \cite{carollo2003geometric, whitney2003berry, tong2004kinematic}.
\par
Soon after the original Berry's work, the GP concept was extended to the generalized coherent states by Klauder et al. \cite{klauder1985coherent, zhang1990coherent}. Adiabatic GP for the coherent states was first studied by Kuratsuji et al. \cite{kuratsuji1985effective}, and subsequently, he and others generalized it to non-adiabatic evolutions \cite{kuratsuji1988geometric, littlejohn1988cyclic}. Since then, noncyclic GPs for coherent and squeezed states have been extensively investigated \cite{mukunda1993quantum,mendas1997pancharatnam,chaturvedi1987berry, pati1995geometric, sjoqvist1997noncyclic, yang2011geometric}. However, almost all of the previous studies are confined to one-mode cases, and little attention has been paid to the study of GP for two-mode coherent and squeezed states.
\par
The concept of GP has found applications in different fields of physics, namely, quantum information and computation science \cite{zanardi1999holonomic, jones2000geometric, zhu2002implementation, vedral2003geometric, rowell2018mathematics}, condensed matter physics\cite{morpurgo1998ensemble, niu1999adiabatic}, and optics \cite{tiwari1992geometric, galvez2002applications}. Due to its purely geometric nature, the GP can be used as a convenient tool for analyzing the characteristics of optical phenomena.  Investigating the GP in entangled quantum systems can help to understand the nature of quantum entanglements \cite{sjoqvist2015geometric}, which has attracted noticeable attention in composite systems. Although GP is studied in detail in entangled spin systems \cite{sjoqvist2000geometric, tong2003geometric, tong2003relation, bertlmann2004berry, najarbashi2017quantum}, investigations for the entangled coherent and squeezed states are scarce.
\par
In this paper, we study the GP in two-mode, entangled quantum states. It is organized as follows: Section two is a brief review on the kinematic approach to the GP for unitary evolutions.
In section 3, we investigate the GP of two-mode entangled squeezed-coherent states (ESCSs), undergoing a unitary cyclic evolution, as well as the effect of the squeezing parameter on it. In section 4, we study the effect of the dimensions of the constituting subsystems on the GP.
In section 5, we suggest a method to generate the two-mode balanced ESCSs, physically, based on an interferometry setup. Finally, the summary and conclusions are given in section 6.

\section{Quantum kinematic approach to GP}
\par
Considering a pure normalized state $|\Psi(t)\rangle$ that evolves from $t=0$ to $t=\tau$ on a path $C$ in the Hilbert space, the dynamical phase is defined as
\begin{equation}\label{dyn}
\Phi_{Dyn}=-i\int\limits_{0}^{\tau} dt \langle\Psi(t)|\dot{\Psi}(t)\rangle,
\end{equation}
with $\langle\Psi(0)|\Psi(\tau)\rangle\neq0$. Based on the definition, $\Phi_{Dyn}$ is obviously energy dependent. Accordingly, the GP associated with the projective Hilbert space is defined as the residue of the dynamical phase, subtracted from the total acquired phase \cite{mukunda1993quantum}:
\begin{equation}\label{gp}
\Phi_{G}[\mathcal{C}]=arg\langle\Psi(0)|\Psi(\tau)\rangle+i\int\limits_{0}^{\tau} dt \langle\Psi(t)|\dot{\Psi}(t)\rangle.
\end{equation}
So, the GP is a property of the path $\mathcal{C}$ in the projective Hilbert space. It is gauge and reparametrization invariant \cite{mukunda1993quantum}. For cyclic evolutions and in the adiabatic limit, $\Phi_{G}[\mathcal{C}]$ reduces to Berry's phase \cite{berry1984quantal}.
One of the most important generalizations of Berry's phase is propounded by Aharanov et al. \cite{aharonov1987phase}. They showed that Berry's phase reduces to the Aharonov-Anandan phase by removing the adiabatic condition. In practice, this quantity is very important because the adiabatic condition is not exactly fulfilled in actual processes.

\section{GP of two-mode ESCSs}
\par
We start this section by defining squeezed-coherent states and reviewing some of their fundamental properties. Basically, general squeezed states can be generated by applying the squeezing operator, $\hat{S}(\xi)$, and the displacement operator, $\hat{D}(\alpha)$, to the vacuum state, respectively:
\begin{equation}
|\alpha,\xi\rangle=\hat{D}(\alpha)\hat{S}(\xi)|0\rangle.
\end{equation}
The squeezing operator is given by $\exp[\frac{1}{2}(\xi^{\ast} \hat{a}^{2}-\xi\hat{a}^{\dag 2})]$, where $\xi=r e^{i \Theta}$, $0\leq r<\infty$ and $0\leq \Theta<2\pi$. The parameter $r$ is called the squeezing parameter. The displacement operator is also defined as $\exp(\alpha \hat{a}^{\dag}-\alpha^{*}\hat{a})$, where $\hat{a}^{\dag}$ and $\hat{a}$ are the bosonic creation and annihilation operators, respectively. Here, $\alpha$ is a complex number which is called the coherence parameter. For $\alpha=0$, the state $|\alpha,\xi\rangle$ is just the squeezed state and for $\xi=0$, it is the coherent state; Hence in general, the state $|\alpha,\xi\rangle$ is called a squeezed-coherent state.
\par
Squeezed-coherent states can also be defined as the eigenstates of the operator $\hat{A}=\hat{a}\cosh r+\hat{a}^{\dag}e^{i \Theta}\sinh r$, i.e.\cite{gerry2005introductory}:
\begin{equation}\label{eigenstate-equation}
\hat{A}|\alpha,\xi\rangle=\eta|\alpha,\xi\rangle,
\end{equation}
where $\eta=\alpha\cosh r+\alpha^{*}e^{i \Theta}\sinh r$. When $\alpha$ and $\xi$ are real numbers, $\eta$  reduces to $\eta=\alpha e^{r}$.

\par
The squeezed-coherent state can be expanded into the occupation number states, ${|n\rangle}$, in the Fock representation as \cite{gerry2005introductory}:
\begin{align}
|\alpha,\xi\rangle=&\frac{1}{\sqrt{\cosh r}}exp[-\frac{1}{2}|\alpha|^{2}-\frac{1}{2}\alpha^{\ast2}e^{i\Theta}\tanh r] \\ \nonumber
&\times \sum\limits_{n=0}^\infty \frac{(\frac{1}{2}e^{i\Theta}\tanh r)^{n/2}}{\sqrt{n!}}H_{n}[\eta(e^{i\Theta}\sinh(2r))^{-1/2}]|n\rangle,
\end{align}
where $H_{n}$s are the Hermite polynomials. The scalar product of two squeezed-coherent states $|\alpha_{0},\xi_{0}\rangle$ and $|\alpha_{1},\xi_{1}\rangle$ can be calculated as:
\begin{align}
\langle\alpha_{0},\xi_{0}|\alpha_{1},\xi_{1}\rangle=&\frac{1}{\sqrt{\cosh r_{0}\cosh r_{1}}}\exp[-\frac{1}{2}\alpha_{0}^{2}(1+\tanh r_{0})-\frac{1}{2}\alpha_{1}^{2}(1+\tanh r_{1})] \\ \nonumber
&\times\sum\limits_{n=0}^{\infty}\frac{1}{2^{n}n!} (\tanh r_{0}\tanh r_{1})^{n/2} H_{n}[\alpha_{0}e^{r_{0}}(\sinh(2r_{0}))^{-1/2}]H_{n}[\alpha_{1}e^{r_{1}}(\sinh(2r_{1}))^{-1/2}],
\end{align}
where, we have assumed that the parameters $\alpha$ and $\xi$ are real.
Using the Mehler's formula \cite{beals2016special}, we have:
\begin{align}
\sum\limits_{n=0}^{\infty}\frac{H_{n}[x]H_{n}[y]s^{n}}{2^{n}n!}=\frac{1}{\sqrt{1-s^{2}}}\exp[\frac{2xys-x^{2}s^{2}-y^{2}s^{2}}{1-s^{2}}].
\end{align}
With the condition $|s|<1$, the scalar product of two squeezed-coherent states can be obtained as:
\begin{align}\label{scalarproduct}
\langle\alpha_{0},\xi_{0}|\alpha_{1},\xi_{1}\rangle=&\frac{1}{\sqrt{\cosh (r_{0}-r_{1})}}\exp[-\frac{1}{2}\alpha_{0}^{2}(1+\tanh r_{0})-\frac{1}{2}\alpha_{1}^{2}(1+\tanh r_{1})] \\ \nonumber
&\times\exp[\frac{\alpha_{0}\alpha_{1}e^{(r_{0}+r_{1})}}{\cosh (r_{0}-r_{1})}-\frac{\alpha_{0}^{2}e^{2r_{0}}\sinh r_{1}}{2\cosh r_{0}\cosh (r_{0}-r_{1})}-\frac{\alpha_{1}^{2}e^{2r_{1}}\sinh r_{0}}{2\cosh r_{1}\cosh (r_{0}-r_{1})}],
\end{align}
where $0<\tanh r_{0}\tanh r_{1}<1$.

\subsection{squeezed-coherent vacuum state}
\par
Let's consider an initial squeezed-coherent vacuum state, for which the first and the second modes of the field are composed of squeezed-coherent and vacuum states, respectively, i.e.:
\begin{equation}\label{ESCS}
|\phi(0)\rangle=\frac{1}{\sqrt{N}}(|\alpha_{0},\xi_{0}\rangle|0\rangle+|\alpha_{1},\xi_{1}\rangle|0\rangle),
\end{equation}
where, $N=2+2p_{01}$ is the normalization factor, with $p_{01}=\langle\alpha_{0},\xi_{0}|\alpha_{1},\xi_{1}\rangle$. Without loss of generality, we assume all the parameters are real numbers. Then, $p_{01}$ is given by  Eq. (\ref{scalarproduct}).
\par
Now, we aim to calculate the GP of the introduced state, undergoing a unitary cyclic evolution. First, we notice that the operator $\hat{A}$ in Eq. (\ref{eigenstate-equation}) satisfies the commutation relation $[\hat{A},\hat{A}^{\dag}]=1$. Hence, we use the Jordan-Schwinger realization of $SU(2)$ algebra with two sets of boson operators $\{\hat{A},\hat{A}^{\dag}\}$ and $\{\hat{B},\hat{B}^{\dag}\}$ respectively, for the first and for the second modes, satisfying the commutation relations $[\hat{A},\hat{A}^{\dag}]=1$, $[\hat{B},\hat{B}^{\dag}]=1$ and $[\hat{A},\hat{B}]=0$. One can construct the Hermitian operators
\begin{align}
\begin{array}{c}
{\hat{J_{x}}=\frac{1}{2}(\hat{A}^{\dag}\hat{B}+\hat{A}\hat{B}^{\dag})},\\ {\hat{J_{y}}=\frac{1}{2i}(\hat{A}^{\dag}\hat{B}-\hat{A}\hat{B}^{\dag})},\\ {\hat{J_{z}}=\frac{1}{2}(\hat{A}^{\dag}\hat{A}-\hat{B}^{\dag}\hat{B})}, \\
\end{array}
\end{align}
which satisfy the angular momentum commutation relations $[\hat{J}_{i}, \hat{J}_{j}]=\varepsilon_{ijk}i\hbar \hat{J}_{k}$, where $\varepsilon_{ijk}$ is the Levi-Civita tensor. Tacking $\hat{J}=\{\hat{J_{x}},\hat{J_{y}},\hat{J_{z}}\}$, for the evolution of the states we can use the rotation operator $e^{-i \phi \hat{J}.\hat{n}}$, which applies a rotation around the axis $\hat{n}$ by the angle $\phi$.
\par
We can define the unitary operator
\begin{equation}\label{unitaryevolution}
\hat{U} (\theta,\varphi)=e^{-i\varphi\hat{J_{z}}}e^{-i\theta\hat{J_{y}}},
\end{equation}
that performs two rotations; first a rotation around the $y$-axis by the angle $\theta$ and then a rotation around the $z$-axis by the angle $\varphi$. Notice that $\hat{U} (\theta,\varphi)$ is a non-local unitary evolution.
\par
Applying $\hat{U}(\theta,\varphi)$ on the operators $\hat{A}$ and $\hat{B}$ gives:
\begin{align}\label{transformation}
\begin{array}{c}
{\hat{U}^{\dag}\hat{A}\hat{U}= e^{-i\frac{\varphi}{2}}(\hat{A}\cos\frac{\theta}{2}-\hat{B}\sin\frac{\theta}{2})}, \\ {\hat{U}^{\dag}\hat{B}\hat{U}= e^{i\frac{\varphi}{2}}(\hat{B}\cos\frac{\theta}{2}+\hat{A}\sin\frac{\theta}{2})}, \\
\end{array}
\end{align}
where we have used the Baker-Campbell-Hausdorff formula. Then, by applying $\hat{U}(\theta,\varphi)$ on $|\phi(0)\rangle$, we get the evolved state as:
\begin{equation}\label{evolved-ESCS}
|\phi(\theta,\varphi)\rangle=\frac{1}{\sqrt{N}}(|(e^{-i\frac{\varphi}{2}}\alpha_{0}\cos\frac{\theta}{2}), r_{0}\rangle|(e^{i\frac{\varphi}{2}}\alpha_{0}\sin\frac{\theta}{2}),r_{0}\rangle+|(e^{-i\frac{\varphi}{2}}\alpha_{1}\cos\frac{\theta}{2}), r_{1}\rangle|(e^{i\frac{\varphi}{2}}\alpha_{1}\sin\frac{\theta}{2}), r_{1}\rangle).
\end{equation}
The overlap of the initial and the final states is positive and real-valued for a fixed $\theta$ and cyclic evolution of $\varphi$ from $0$ to $2\pi$, and the total phase $\Phi_{total}=arg(\langle\psi(0)|\psi(\theta,\varphi=2\pi)\rangle)$ vanishes, consequently. Thus, according to Eq. (\ref{gp}), we have $\Phi_{G}=-\Phi_{Dyn}$. 
\par
To obtain the GP, we calculate the expectation value of $\hat{U}^{\dag}(\theta,\varphi)\partial_{\varphi}\hat{U}(\theta,\varphi)$ for the initial state $|\phi(0)\rangle$:
\begin{equation}\label{x-ESCS}
\langle\phi(0)|\hat{U}^{\dag}(\theta,\varphi)\partial_{\varphi}\hat{U}(\theta,\varphi)|\phi(0)\rangle=-i \cos\theta \langle\phi(0)|\hat{J_{z}}|\phi(0)\rangle.
\end{equation}
Note that $\langle\phi(0)|\hat{J_{x}}|\phi(0)\rangle=0$ and $\langle\phi(0)|\hat{J_{z}}|\phi(0)\rangle$ is calculated as follows:
\begin{equation}
\langle\phi(0)|\hat{J_{z}}|\phi(0)\rangle=\frac{1}{2N}(\eta_{0}^{2}+\eta_{1}^{2}+2p_{01}\eta_{0}\eta_{1}).
\end{equation}
Putting Eq.(\ref{x-ESCS}) in Eq.(\ref{gp}) and changing $\varphi$ from $0$ to $2\pi$, for a fixed value of $\theta$ the GP is obtained as:
\begin{equation}\label{gp-ESCS}
\Phi_{G}(|\phi(0)\rangle)=\frac{\pi \cos\theta}{N}(\eta_{0}^{2}+\eta_{1}^{2}+2p_{01}\eta_{0}\eta_{1}),
\end{equation}
which depends on $\theta$ and the occupation number of the individual modes of the initial state. Note that we have assumed that all the parameters are real so, $\eta_{i}=\alpha_{i} e^{r_{i}}$, with $i=0,1$. 
\par
\begin{figure}
 \centering
 \subfigure[$r_{0}=0$, $r_{1}=0$]{\includegraphics[width=5cm]{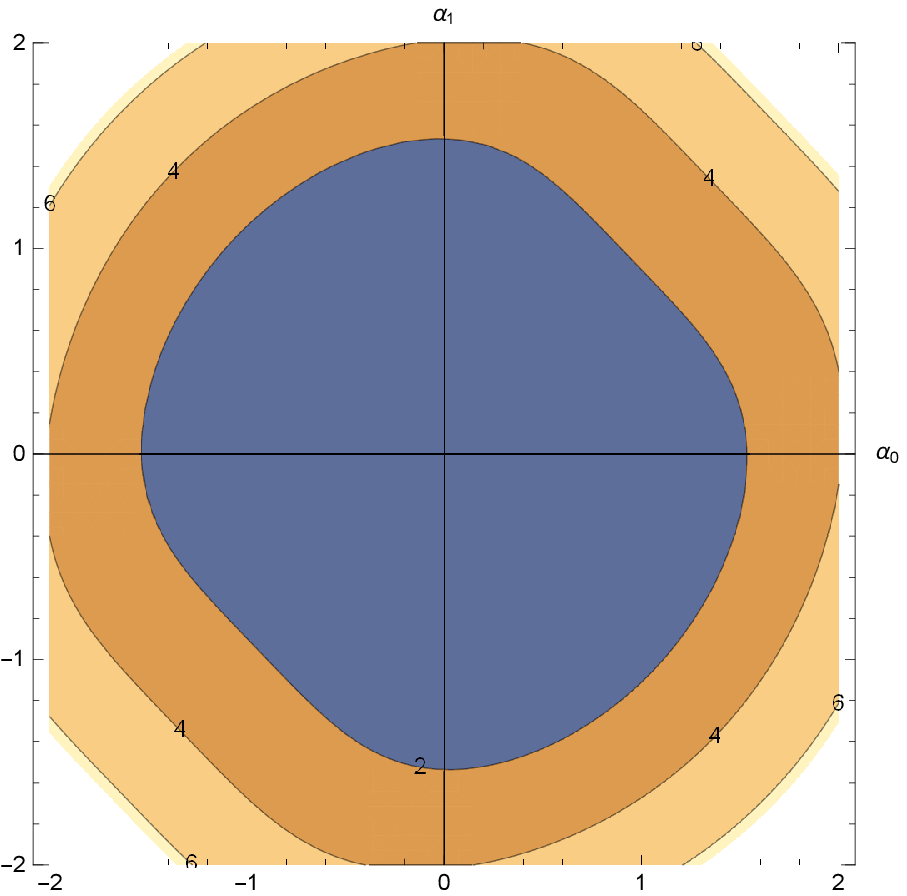}\label{1a}}
  \hfill
 \subfigure[$r_{0}=0.5$, $r_{1}=0.5$]{\includegraphics[width=5cm]{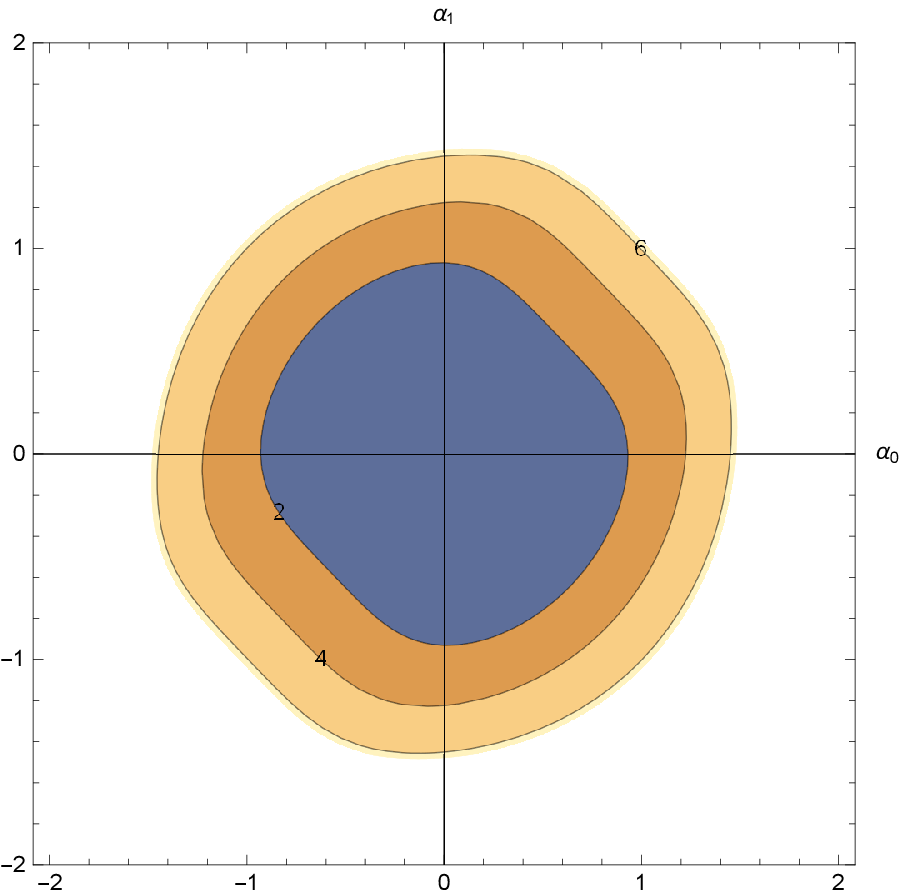}\label{1b}}
  \hfill
 \subfigure[$r_{0}=1$, $r_{1}=1$]{\includegraphics[width=5cm]{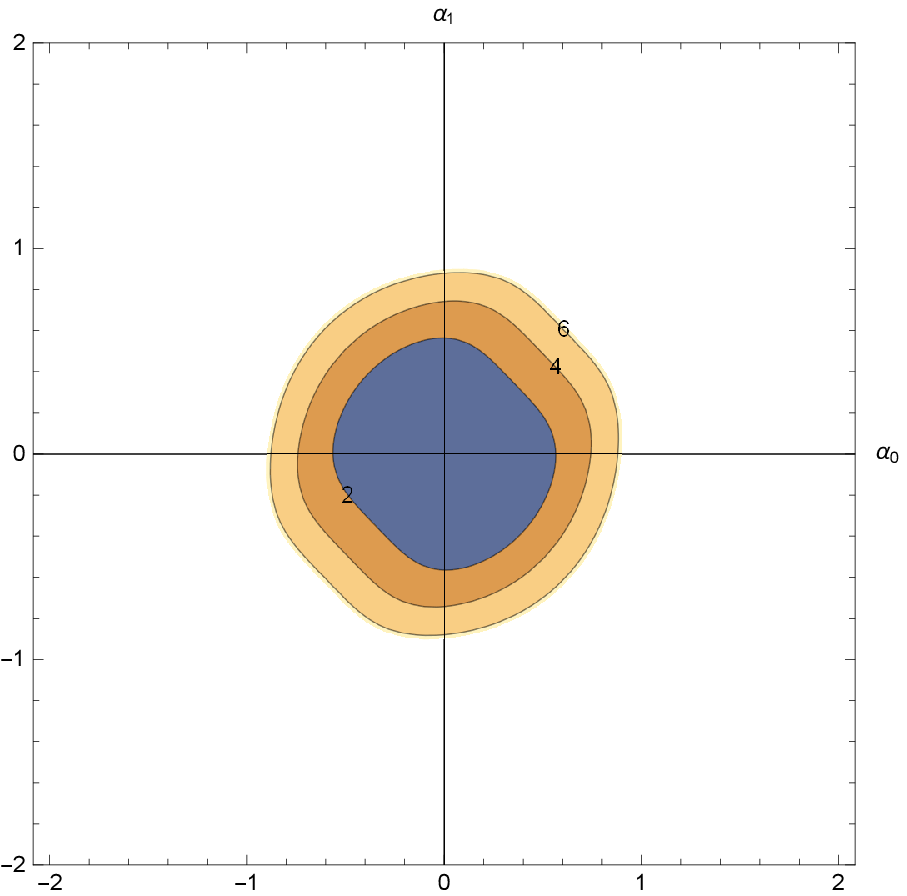}\label{1c}}
  \hfill
 \subfigure[$r_{0}=0$, $r_{1}=0.4$]{\includegraphics[width=5cm]{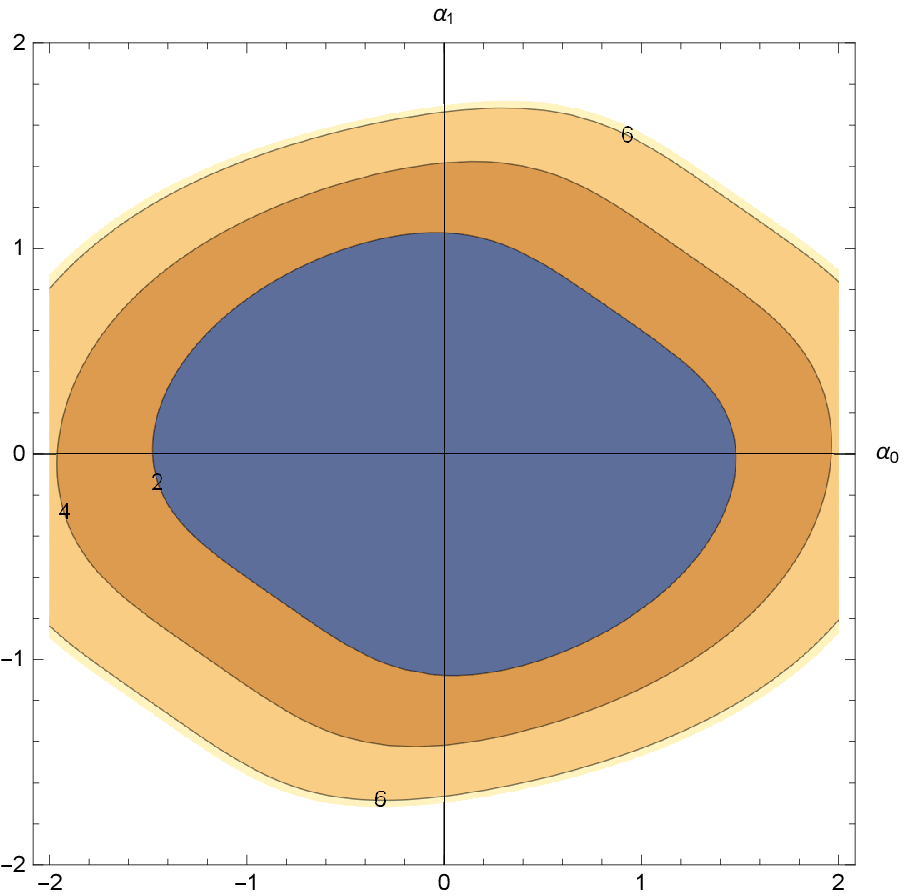}\label{1d}}
 \hfill
  \subfigure[$r_{0}=0$, $r_{1}=0.8$]{\includegraphics[width=5cm]{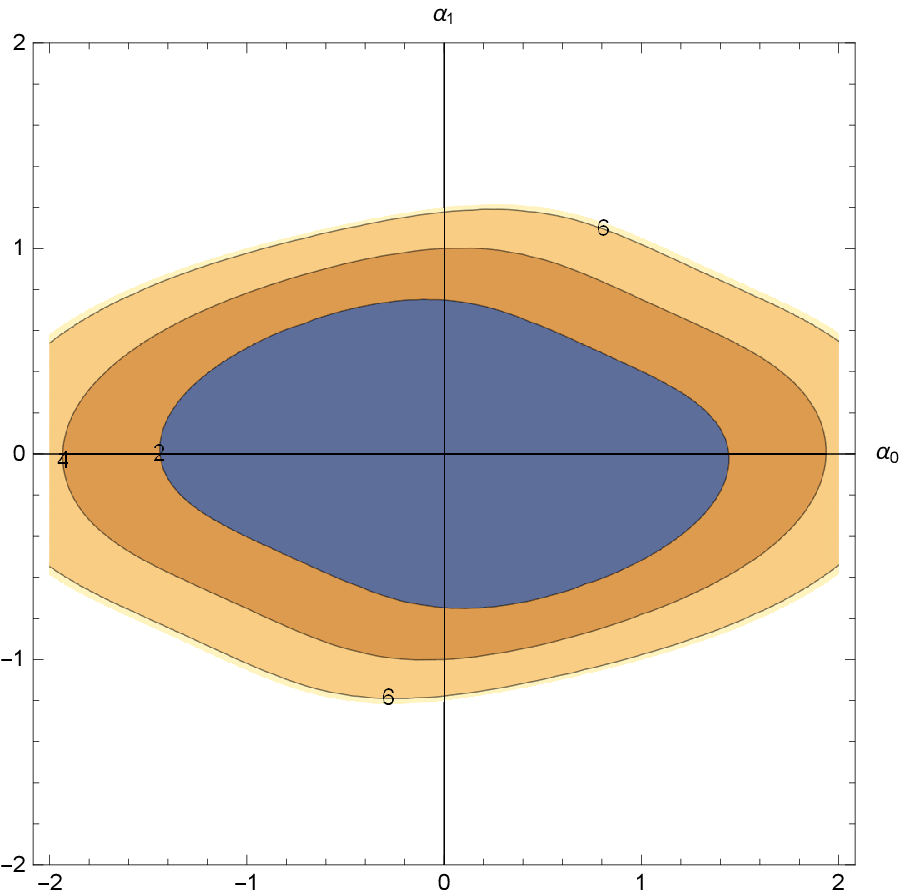}\label{1e}}
 \hfill
  \subfigure[$r_{0}=0$, $r_{1}=1.2$]{\includegraphics[width=5cm]{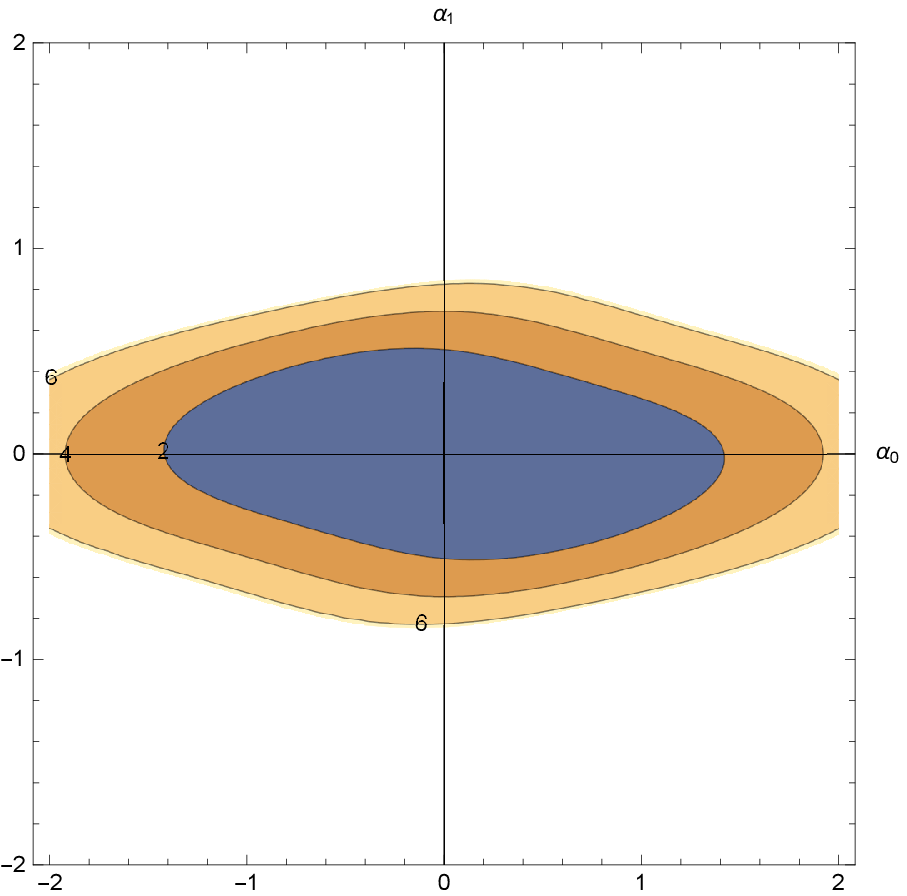}\label{1f}}
 \hfill
  \subfigure[$r_{0}=0.4$, $r_{1}=0$]{\includegraphics[width=5cm]{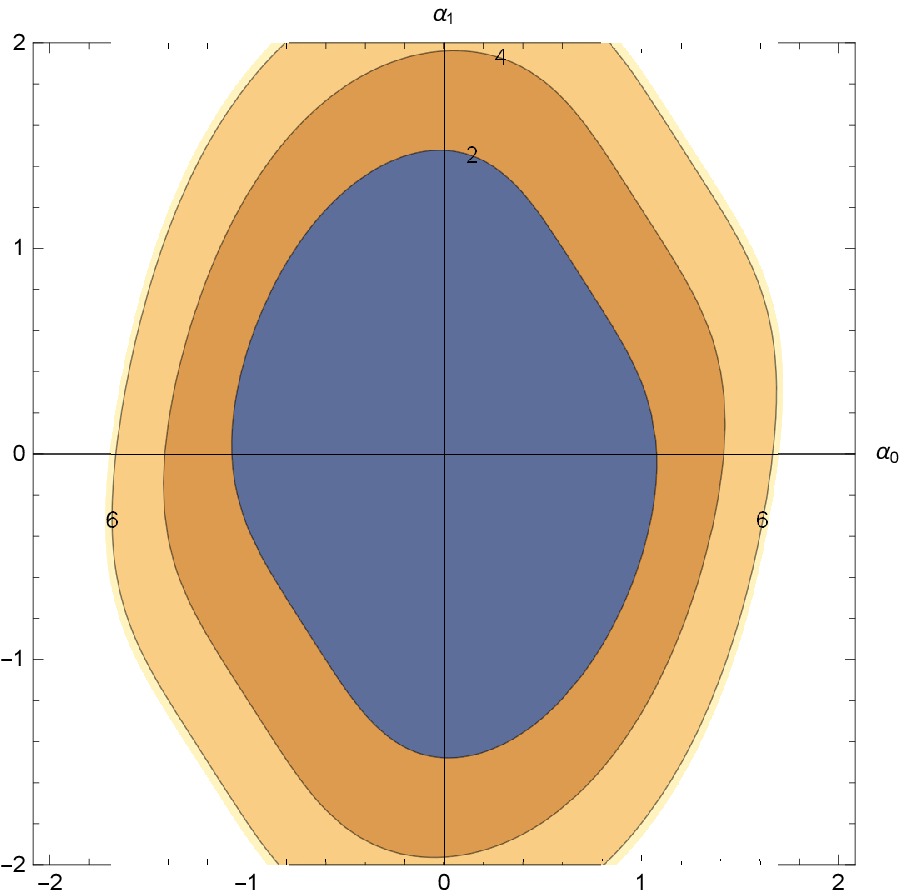}\label{1g}}
 \hfill
  \subfigure[$r_{0}=0.8$, $r_{1}=0$]{\includegraphics[width=5cm]{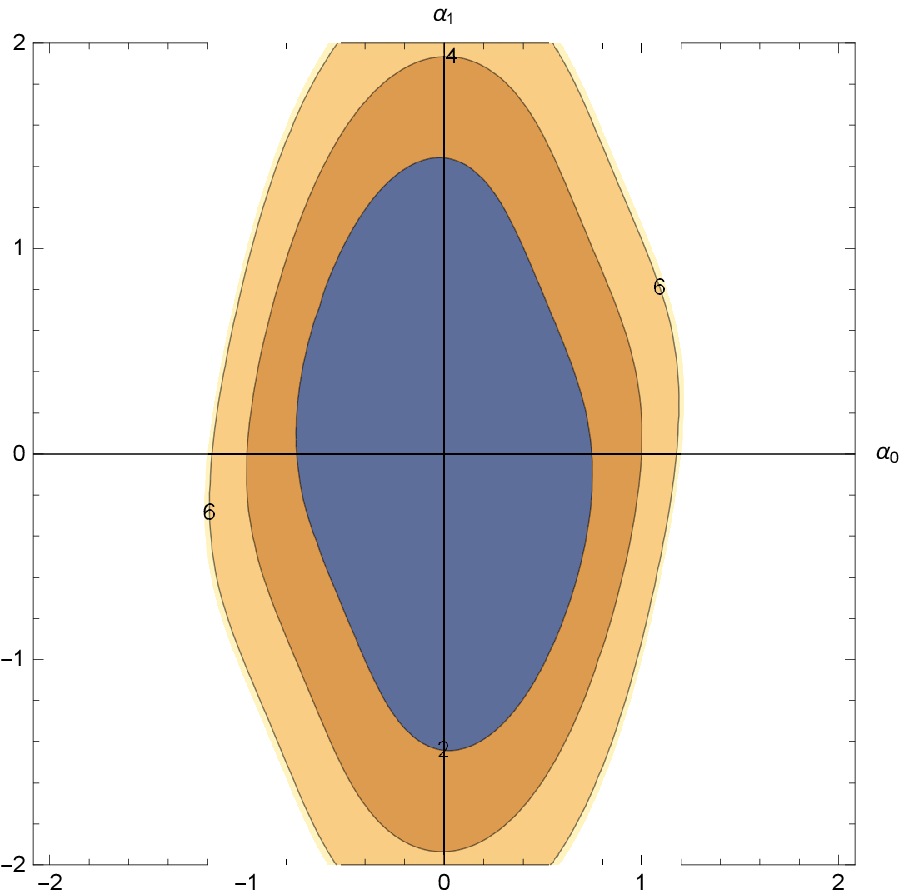}\label{1h}}
 \hfill
  \subfigure[$r_{0}=1.2$, $r_{1}=0$]{\includegraphics[width=5cm]{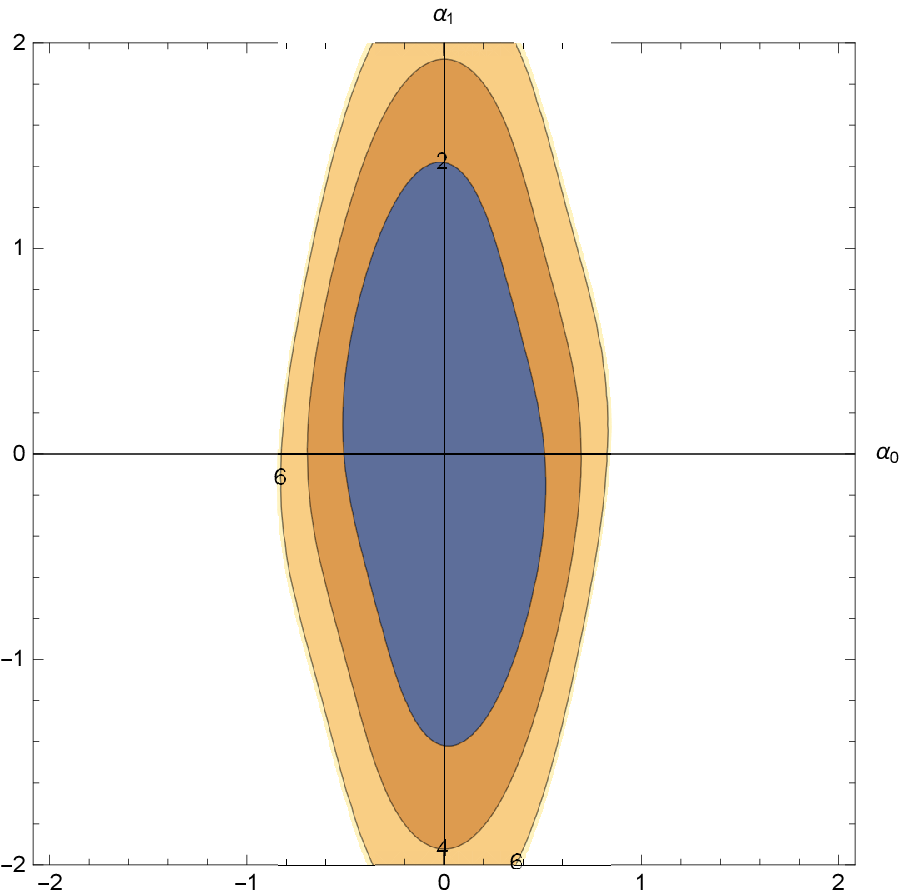}\label{1i}}
  \caption{\small{(Color online.) Contour plots of the GP of $|\phi(0)\rangle$, as functions of $\alpha_{0}$ and $\alpha_{1}$, for $\theta=\pi/4$ and for different values of the squeezing parameters.}}\label{ContourgpESCS}
 \end{figure}

In Fig. \ref{ContourgpESCS}, the calculated GPs for different values of the squeezing parameter are plotted as functions of $\alpha_{0}$ and $\alpha_{1}$. The value of $\theta$ is kept constant and equal to $\pi/4$ in all diagrams. The plots show that GPs are small for small values of the coherence parameters and increase by increasing them. In addition, it is seen that by increasing the squeezing parameters, the GPs are compressed. For example, for a constant value of $r_{0}$($r_{1})$, increasing $r_{1}$($r_{0}$) results in a compression of GP along the axis of $\alpha_{1}$($\alpha_{0}$), in an elliptical manner. 

\subsection{two-mode balanced ESCS}
\par
Now, we consider a two-mode balanced ESCS in the form of:
\begin{equation}\label{balESCS}
|\psi(0)\rangle_{bal}=\frac{1}{\sqrt{M}}(|\alpha_{0},\xi_{0}\rangle|\alpha_{0},\xi_{0}\rangle+|\alpha_{1},\xi_{1}\rangle|\alpha_{1},\xi_{1}\rangle),
\end{equation}
where, $M=2+2p_{01}^{2}$ is the normalization factor, with $p_{01}=\langle\alpha_{0},\xi_{0}|\alpha_{1},\xi_{1}\rangle$.
\par
Using Eq. (\ref{transformation}) and applying $\hat{U}(\theta,\varphi)$ on the initial balanced ESCS yields:
\begin{align}\label{evolved-balESCS}
|\psi(\theta,\varphi)\rangle_{bal}=&\frac{1}{\sqrt{M}}(|(e^{-i\frac{\varphi}{2}}\alpha_{0}(\cos\frac{\theta}{2}-\sin\frac{\theta}{2})), r_{0}\rangle|(e^{i\frac{\varphi}{2}}\alpha_{0}(\cos\frac{\theta}{2}+\sin\frac{\theta}{2})),r_{0}\rangle+\\ \nonumber
&|(e^{-i\frac{\varphi}{2}}\alpha_{1}(\cos\frac{\theta}{2}-\sin\frac{\theta}{2})), r_{1}\rangle|(e^{i\frac{\varphi}{2}}\alpha_{1}(\cos\frac{\theta}{2}+\sin\frac{\theta}{2})), r_{1}\rangle).
\end{align}
It is easy to check that the total phase is zero and to obtain the GP, we just need to calculate the dynamical phase.
The expectation value of $\hat{U}^{\dag}(\theta,\varphi)\partial_{\varphi}\hat{U}(\theta,\varphi)$ for the initial state, $|\psi(0)\rangle_{bal}$, can be calculated as:
\begin{equation}\label{x-bal}
_{bal}\langle\hat{U}^{\dag}(\theta,\varphi)\partial_{\varphi}\hat{U}(\theta,\varphi)\rangle_{bal}=i \frac{\sin\theta}{M} (\eta_{0}^{2}+\eta_{1}^{2}+2p_{01}^{2}\eta_{0}\eta_{1}).
\end{equation}
Then, the cyclic GP for the balanced ESCS takes the form:
\begin{equation}\label{gp-balESCS}
\Phi_{G}(|\psi(0)\rangle_{bal})=\frac{-2\pi \sin\theta}{M}(\eta_{0}^{2}+\eta_{1}^{2}+2p_{01}^{2}\eta_{0}\eta_{1}).
\end{equation}
For $r=0$, it reduces to the GP for the balanced entangled coherent state (ECS) \cite{Almas2022geometric}.

\par
\begin{figure}
 \centering
 \subfigure[$r_{0}=0$, $r_{1}=0$]{\includegraphics[width=5cm]{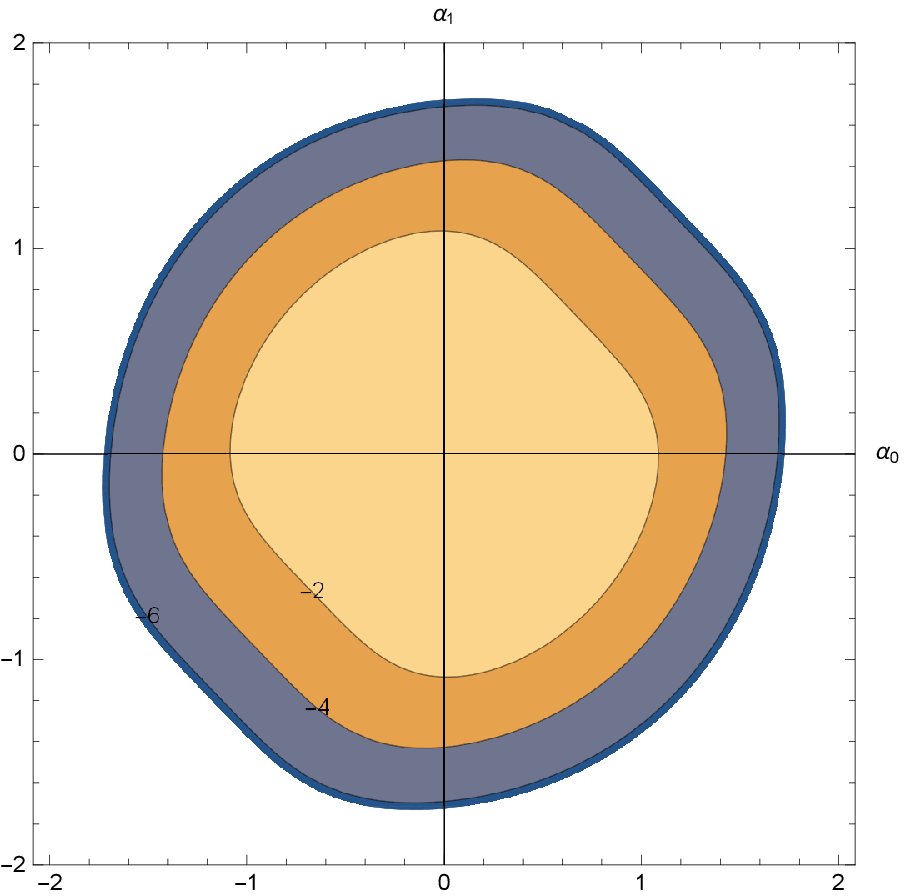}\label{2a}}
  \hfill
 \subfigure[$r_{0}=0.5$, $r_{1}=0.5$]{\includegraphics[width=5cm]{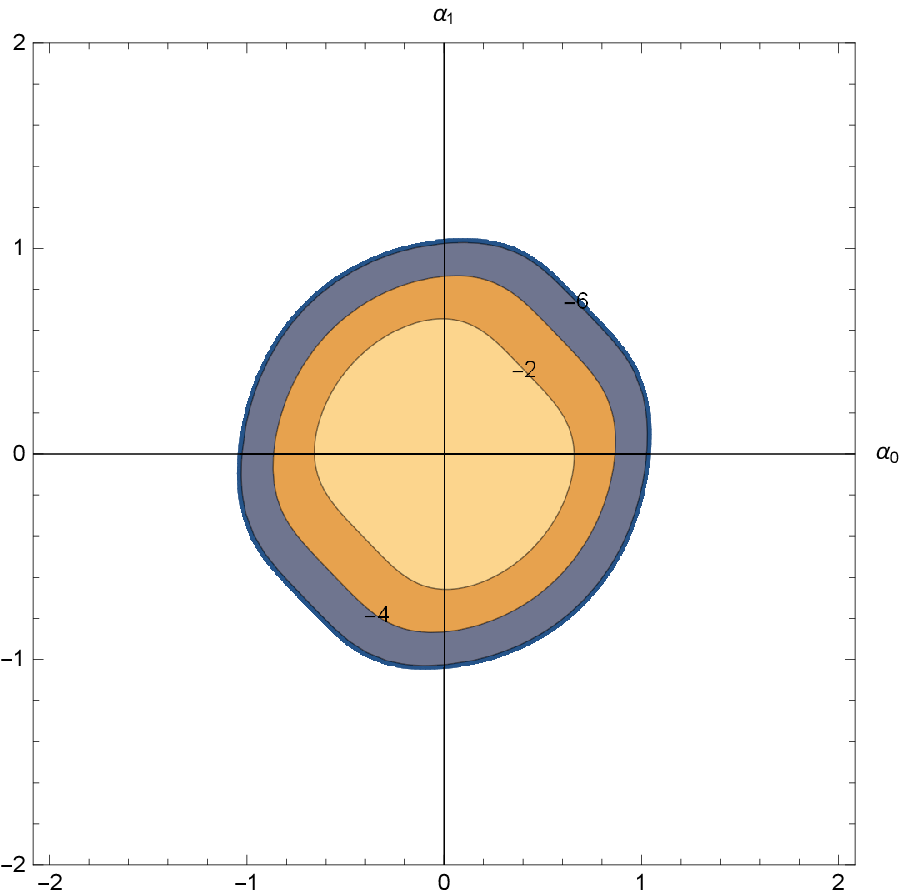}\label{2b}}
  \hfill
 \subfigure[$r_{0}=1$, $r_{1}=1$]{\includegraphics[width=5cm]{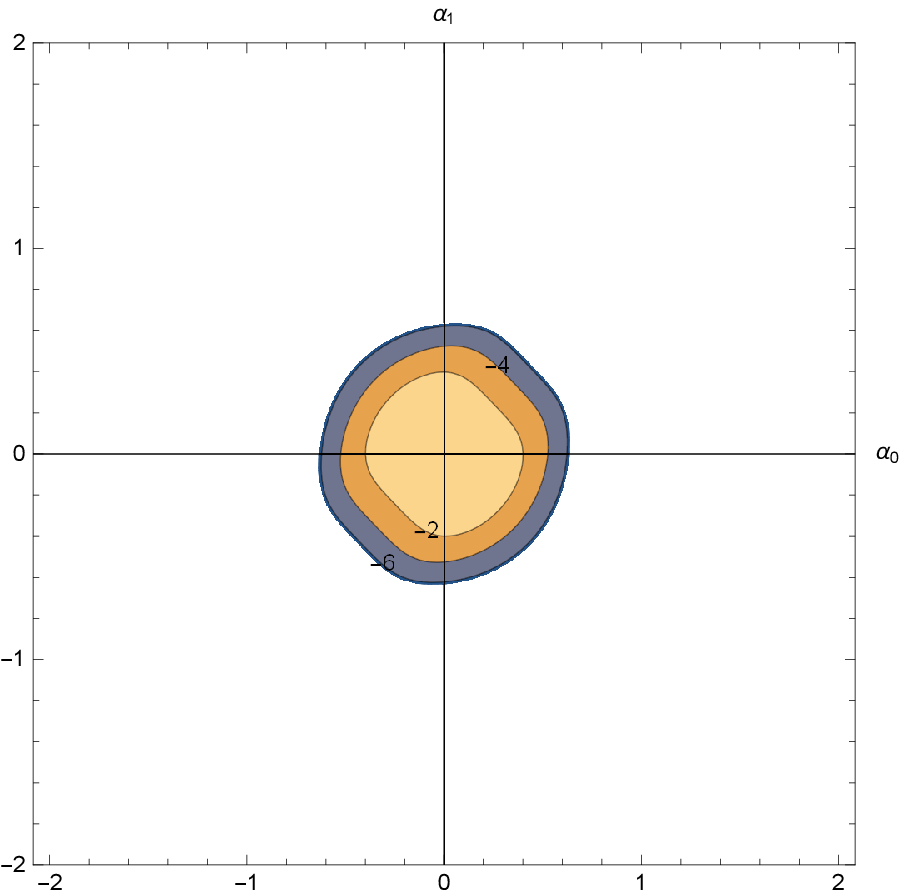}\label{2c}}
  \hfill
 \subfigure[$r_{0}=0$, $r_{1}=0.4$]{\includegraphics[width=5cm]{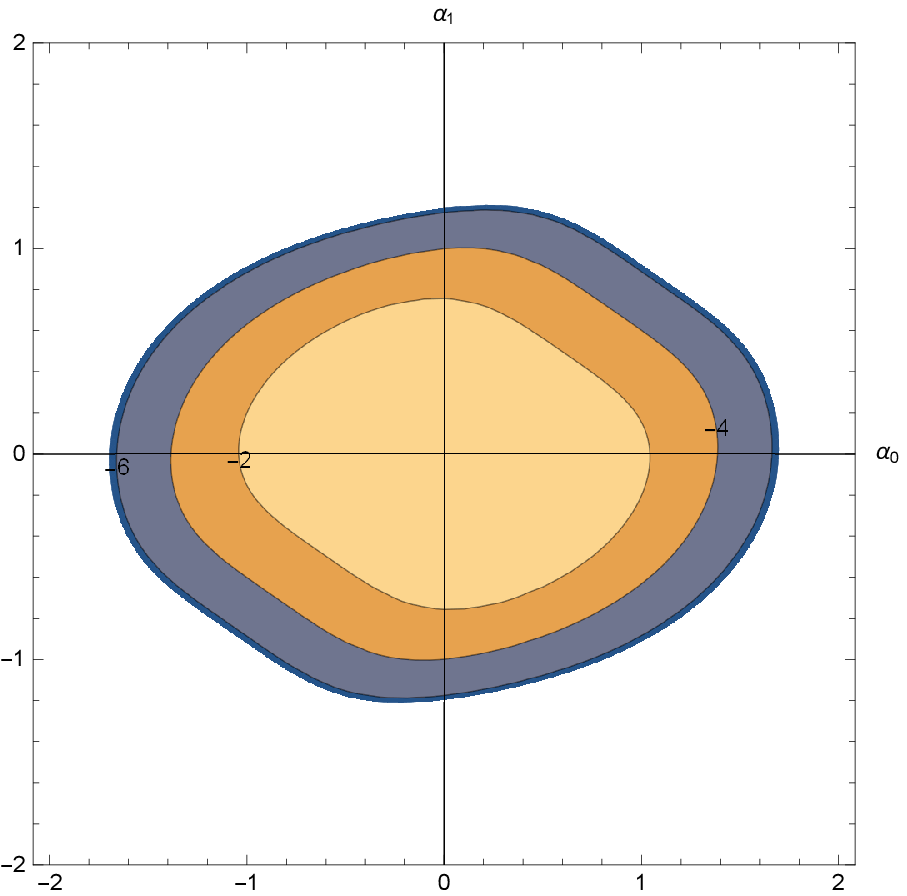}\label{2d}}
 \hfill
  \subfigure[$r_{0}=0$, $r_{1}=0.8$]{\includegraphics[width=5cm]{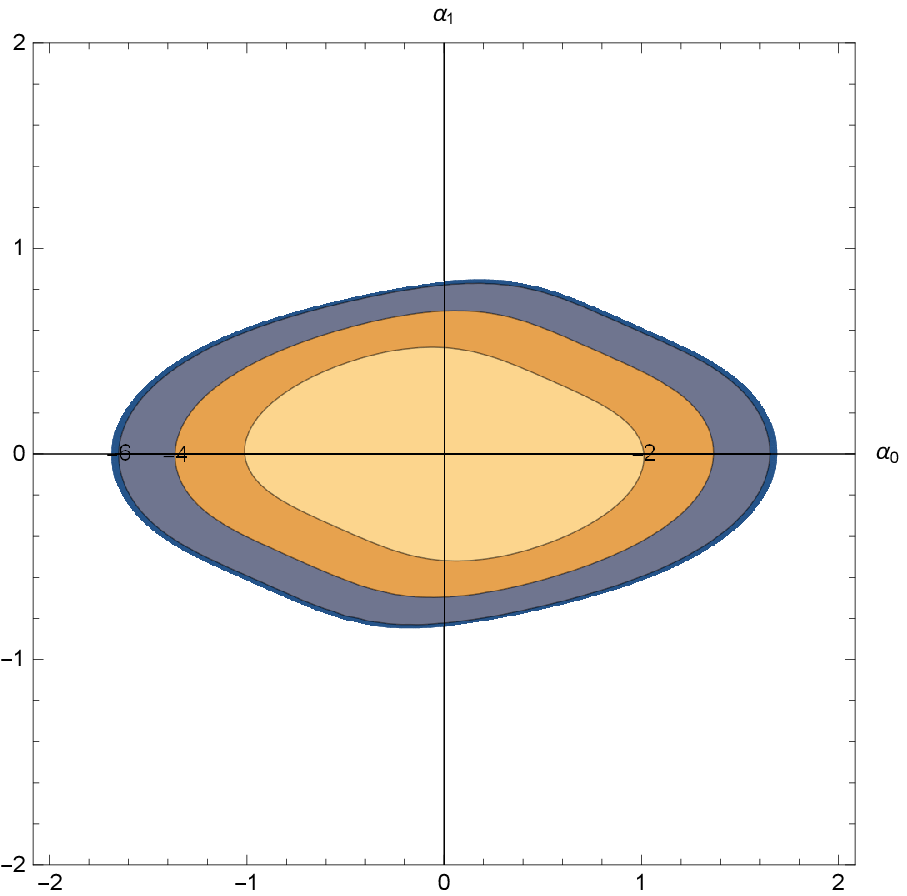}\label{2e}}
 \hfill
  \subfigure[$r_{0}=0$, $r_{1}=1.2$]{\includegraphics[width=5cm]{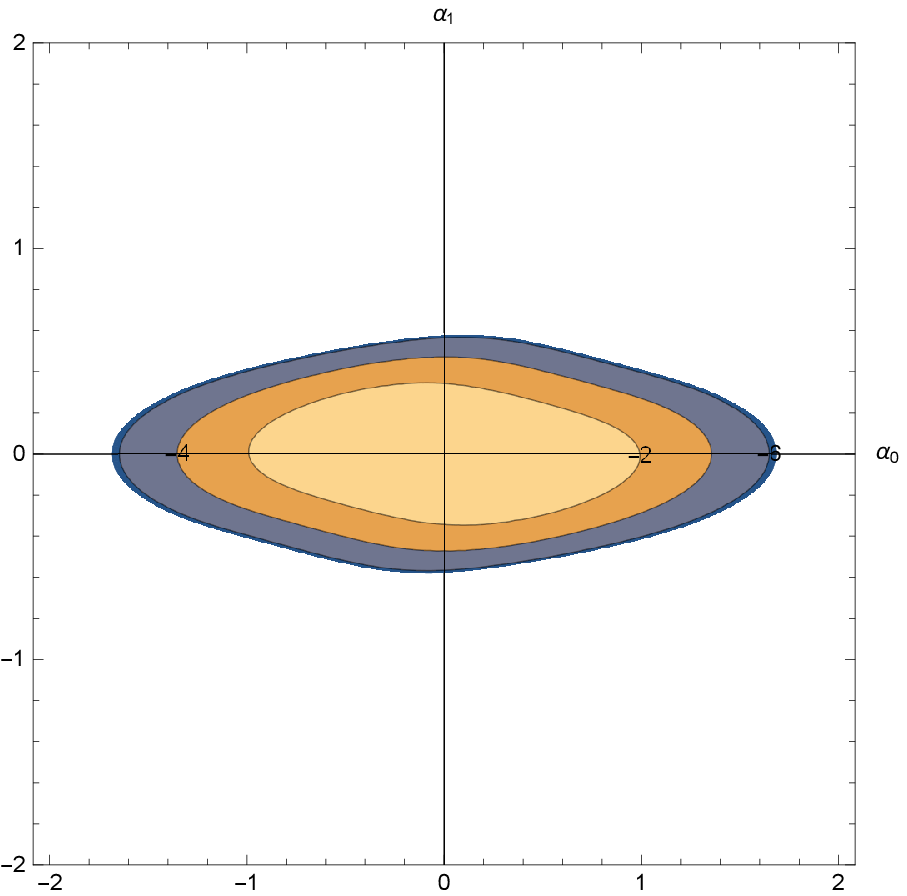}\label{2f}}
 \hfill
  \subfigure[$r_{0}=0.4$, $r_{1}=0$]{\includegraphics[width=5cm]{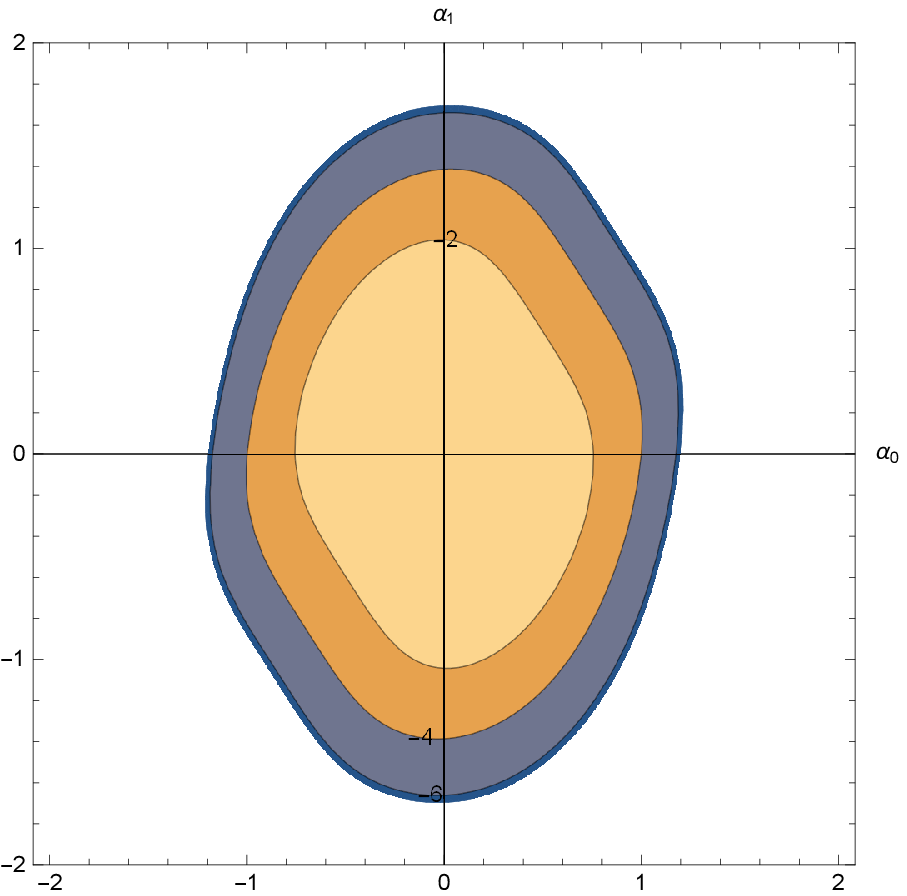}\label{2g}}
 \hfill
  \subfigure[$r_{0}=0.8$, $r_{1}=0$]{\includegraphics[width=5cm]{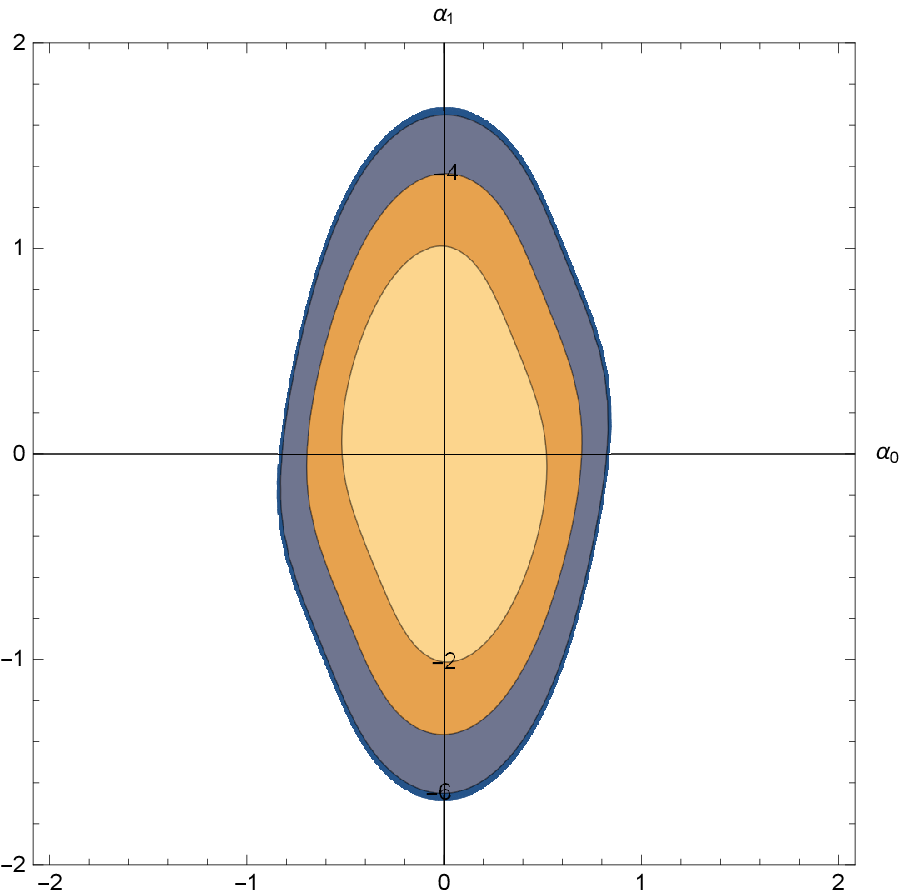}\label{2h}}
 \hfill
  \subfigure[$r_{0}=1.2$, $r_{1}=0$]{\includegraphics[width=5cm]{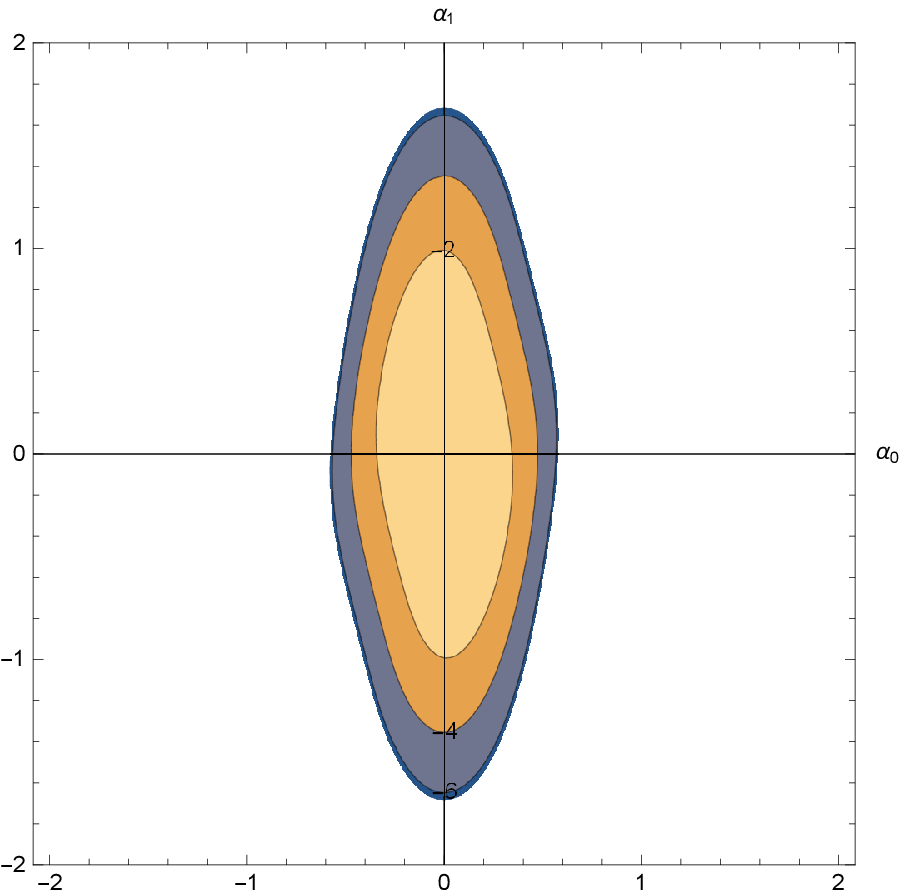}\label{2i}}
  \caption{\small{(Color online.) 
  Contour plots of the GP of $|\phi(0)\rangle_{bal}$, as functions of $\alpha_{0}$ and $\alpha_{1}$, for $\theta=\pi/4$ and for different values of the squeezing parameters.}}\label{ContourgpbalESCS}
 \end{figure}

In Fig. \ref{ContourgpbalESCS}, the calculated GPs for different values of the squeezing parameter are plotted as functions of $\alpha_{0}$ and $\alpha_{1}$, for the $|\psi(0)\rangle_{bal}$ state. The plots are similar to that of the $|\phi(0)\rangle$ state. As the figure shows, for a fixed value of $r_{0}$($r_{1})$, increasing $r_{1}$($r_{0}$) results in a compression of GP along the axis of $\alpha_{1}$($\alpha_{0}$), in an elliptical manner. 

\par
\begin{figure}
 \centering
 \subfigure[$r_{0}=0$]{\includegraphics[width=8cm]{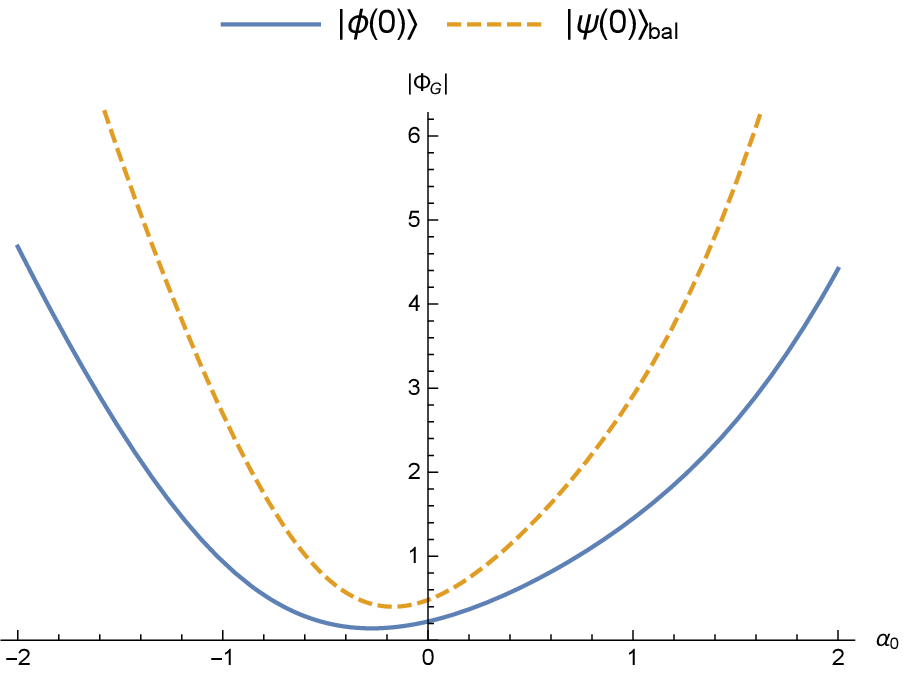}\label{3a}}
  \hfill
  \subfigure[$r_{0}=0.5$]{\includegraphics[width=8cm]{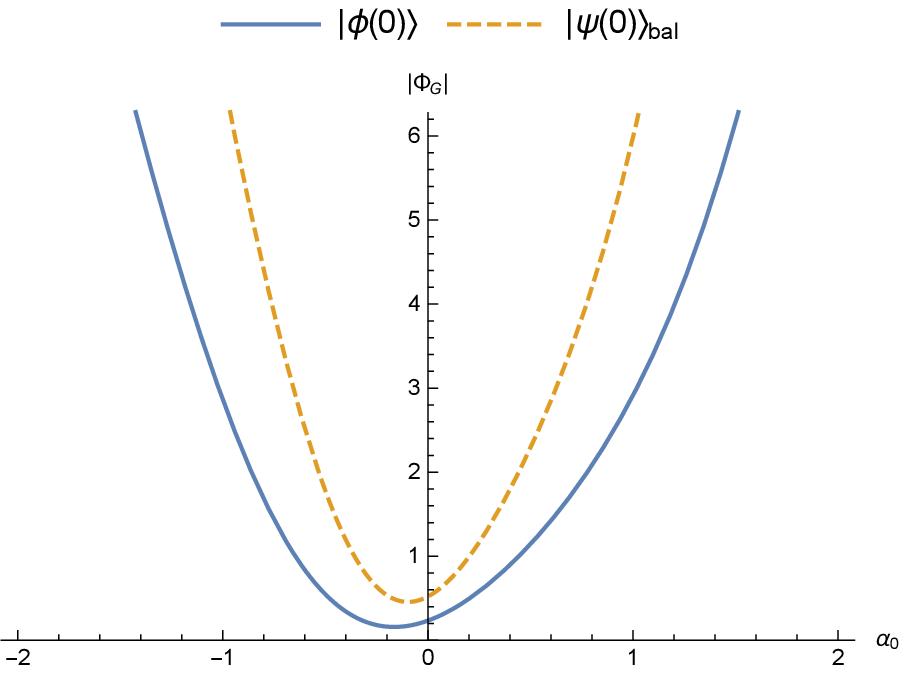}\label{3b}}
 \hfill
  \subfigure[$r_{0}=1$]{\includegraphics[width=8cm]{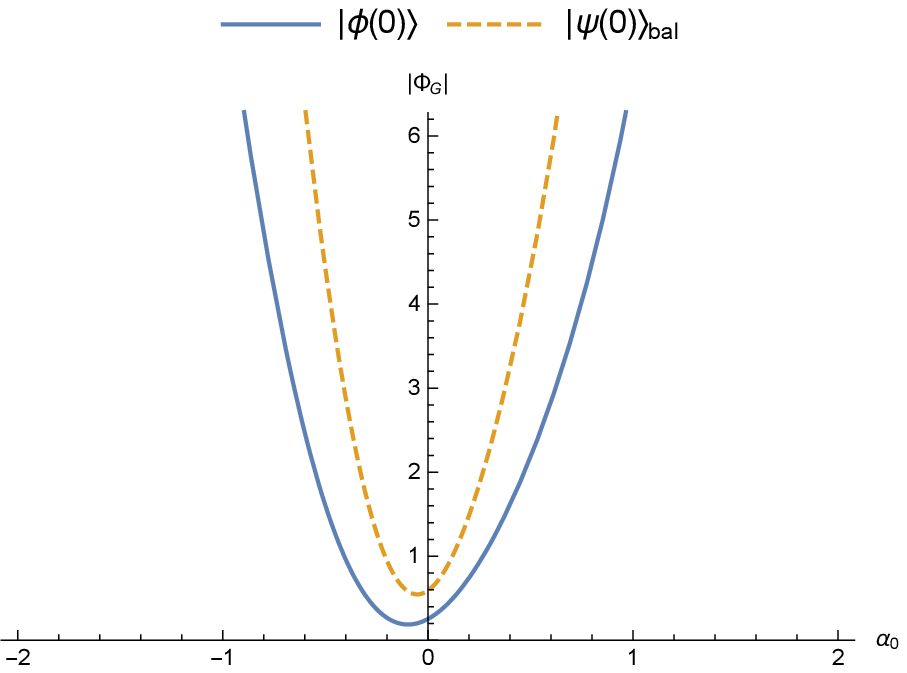}\label{3c}}
 \hfill
  \subfigure[$r_{0}=1.5$]{\includegraphics[width=8cm]{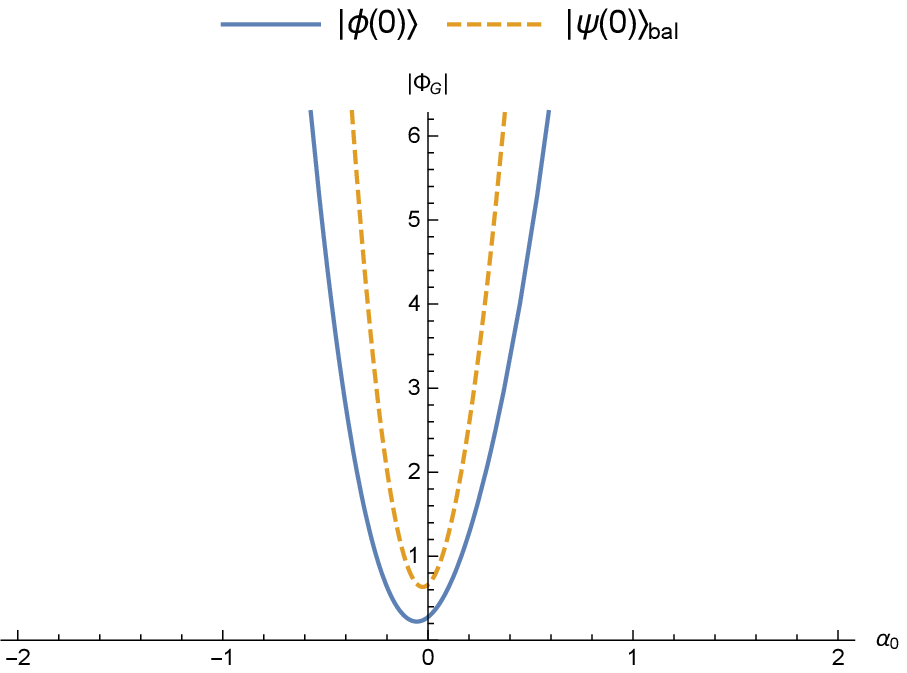}\label{3d}}
  \caption{\small{The modulus of the GPs of the two $|\phi(0)\rangle$ (solid lines) and $|\psi(0)\rangle_{bal}$ (dashed lines) states as functions of $\alpha_{0}$, for $\alpha_{1}=0.5$, $r_{1}=0.2$, $\theta=\pi/4$, and for different values of $r_{0}$.}}\label{comparisonESCS}
 \end{figure}

In Fig. \ref{comparisonESCS}, the GP of two states, $|\phi(0)\rangle$ and $|\psi(0)\rangle_{bal}$ are compared with eachother. The figures show that the GPs of both states have the same behavior, with the only difference that the GP of $|\psi(0)\rangle_{bal}$ increases faster than the GP of $|\phi(0)\rangle$ by increasing $|\alpha_{0}|$.

\subsection{two-mode unbalanced ESCS}
\par
Now, let's consider a two-mode unbalanced ESCS and calculate its GP under the unitary evolution. The state is:
\begin{equation}\label{unbalESCS}
|\psi(0)\rangle_{unbal}=\frac{1}{\sqrt{M}}(|\alpha_{0},\xi_{0}\rangle|\alpha_{1},\xi_{1}\rangle+|\alpha_{1},\xi_{1}\rangle|\alpha_{0},\xi_{0}\rangle),
\end{equation}
where, $M=2+2p_{01}^{2}$ is the normalization factor, with $p_{01}=\langle\alpha_{0},\xi_{0}|\alpha_{1},\xi_{1}\rangle$.
\par
Applying $\hat{U}(\theta,\varphi)$ on the unbalanced ESCS yields:
\begin{align}\label{evolved-unbalESCS}
|\psi(\theta,\varphi)\rangle_{unbal}=&\frac{1}{\sqrt{M}}(|(e^{-i\frac{\varphi}{2}}(\alpha_{0}\cos\frac{\theta}{2}-\alpha_{1}\sin\frac{\theta}{2})), r_{0}\rangle|(e^{i\frac{\varphi}{2}}(\alpha_{1}\cos\frac{\theta}{2}+\alpha_{0}\sin\frac{\theta}{2})),r_{1}\rangle\\ \nonumber
&+|(e^{-i\frac{\varphi}{2}}(\alpha_{1}\cos\frac{\theta}{2}-\alpha_{0}\sin\frac{\theta}{2})), r_{1}\rangle|(e^{i\frac{\varphi}{2}}(\alpha_{0}\cos\frac{\theta}{2}+\alpha_{1}\sin\frac{\theta}{2})), r_{0}\rangle),
\end{align}
where we have used Eq. (\ref{transformation}). 
\par
The expectation value of $\hat{U}^{\dag}(\theta,\varphi)\partial_{\varphi}\hat{U}(\theta,\varphi)$ for the initial state $|\psi(0)\rangle_{unbal}$ can be obtained as:
\begin{equation}\label{x-unbal}
_{unbal}\langle\hat{U}^{\dag}(\theta,\varphi)\partial_{\varphi}\hat{U}(\theta,\varphi)\rangle_{unbal}=i \frac{\sin\theta}{M} ((\eta_{0}^{2}+\eta_{1}^{2})p_{01}^{2}+2\eta_{0}\eta_{1}).
\end{equation}
Then, the cyclic GP for the unbalanced ESCS takes the form:
\begin{equation}\label{gp-unbalESCS}
\Phi_{G}(|\psi(0)\rangle_{unbal})=\frac{-2\pi \sin\theta}{M}((\eta_{0}^{2}+\eta_{1}^{2})p_{01}^{2}+2\eta_{0}\eta_{1}).
\end{equation}
It is worth noting that for $r=0$, the GP reduces to the GP of the unbalanced ECS, introduced in \cite{Almas2022geometric}.
\par
\begin{figure}
 \centering
 \subfigure[$r_{0}=0$, $r_{1}=0$]{\includegraphics[width=5cm]{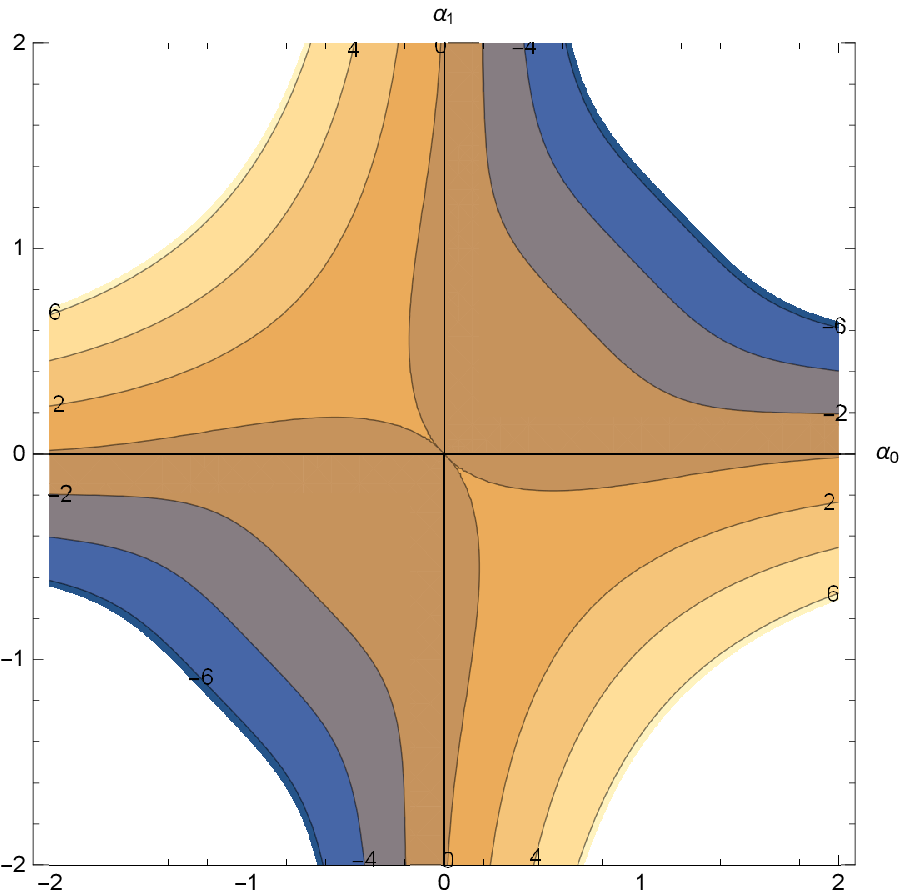}\label{4a}}
  \hfill
 \subfigure[$r_{0}=0.5$, $r_{1}=0.5$]{\includegraphics[width=5cm]{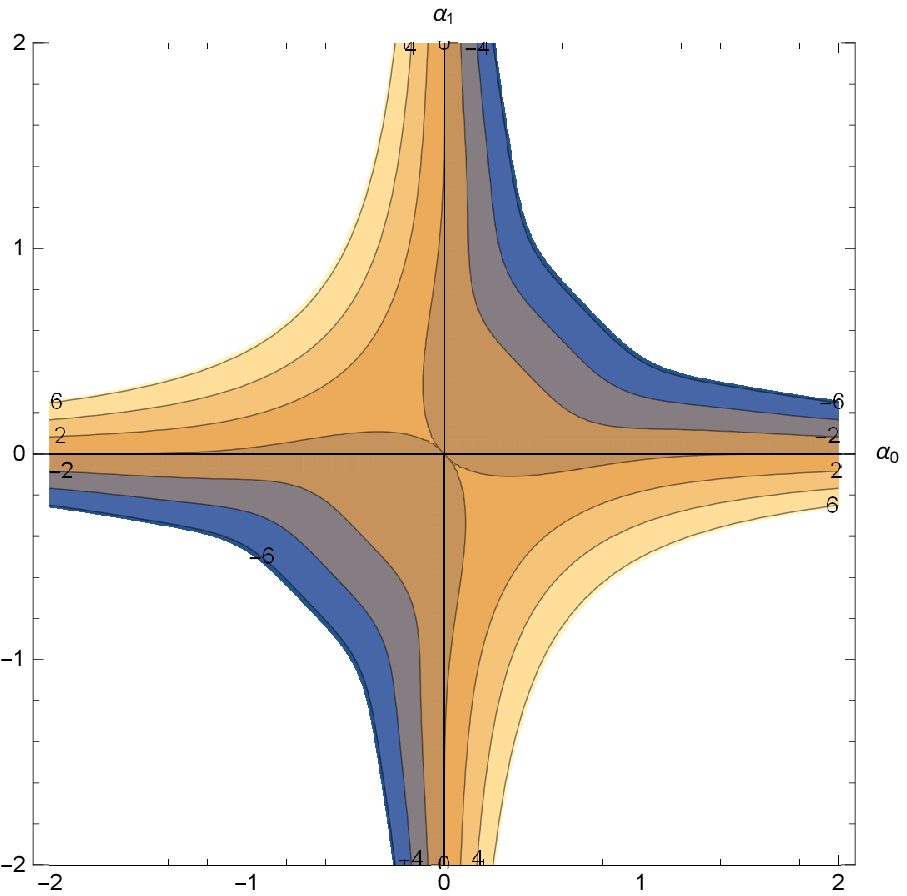}\label{4b}}
  \hfill
 \subfigure[$r_{0}=1$, $r_{1}=1$]{\includegraphics[width=5cm]{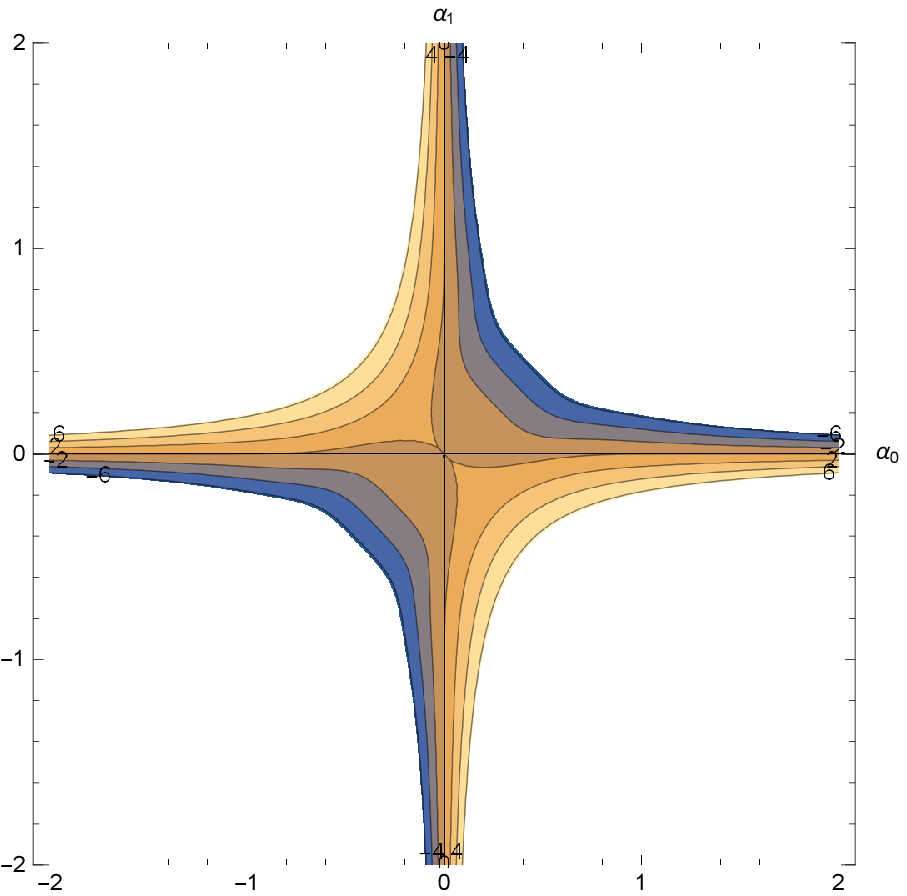}\label{4c}}
  \hfill
 \subfigure[$r_{0}=0$, $r_{1}=0.4$]{\includegraphics[width=5cm]{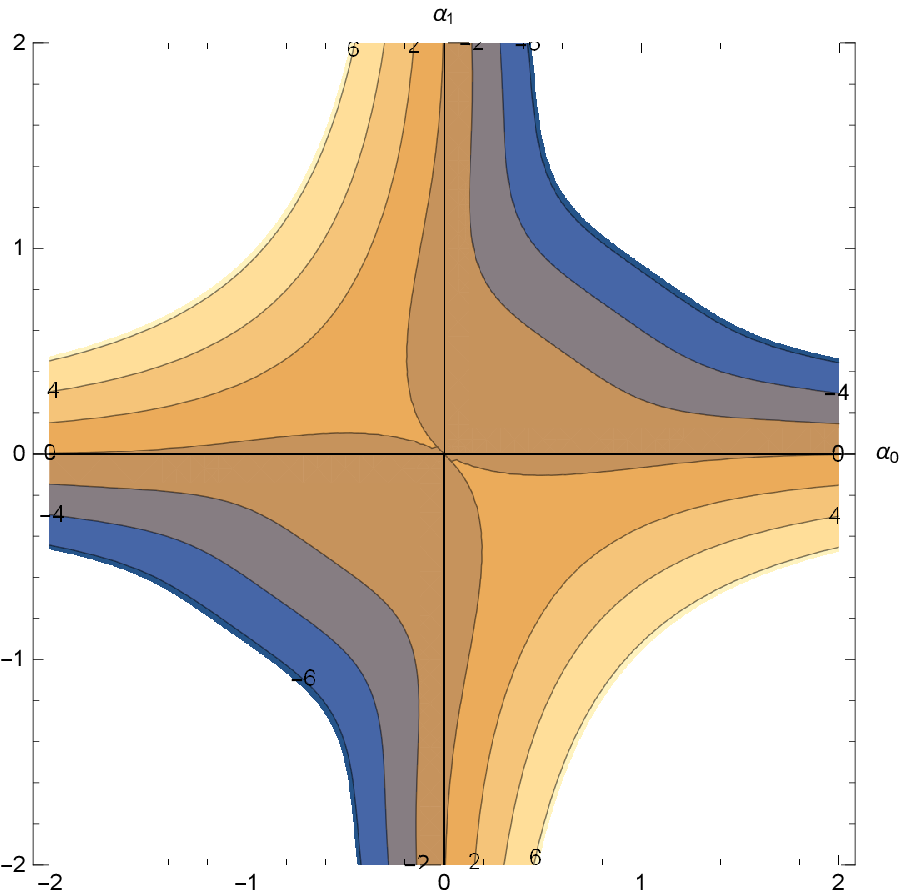}\label{4d}}
 \hfill
  \subfigure[$r_{0}=0$, $r_{1}=0.8$]{\includegraphics[width=5cm]{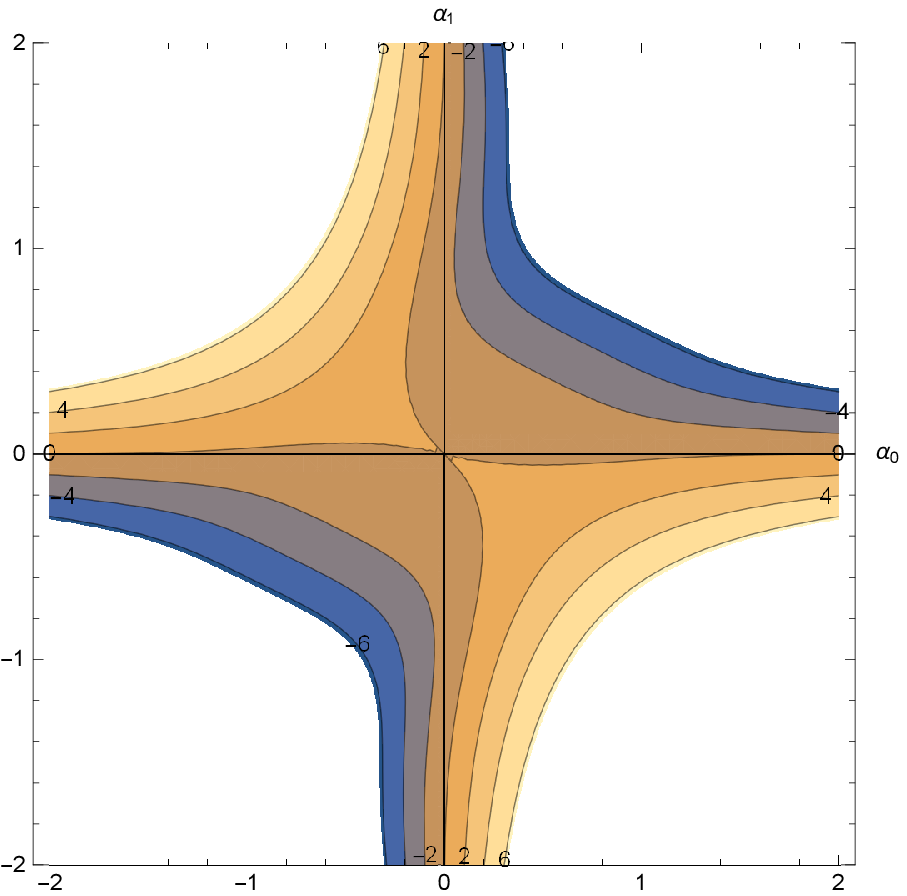}\label{4e}}
 \hfill
  \subfigure[$r_{0}=0$, $r_{1}=1.2$]{\includegraphics[width=5cm]{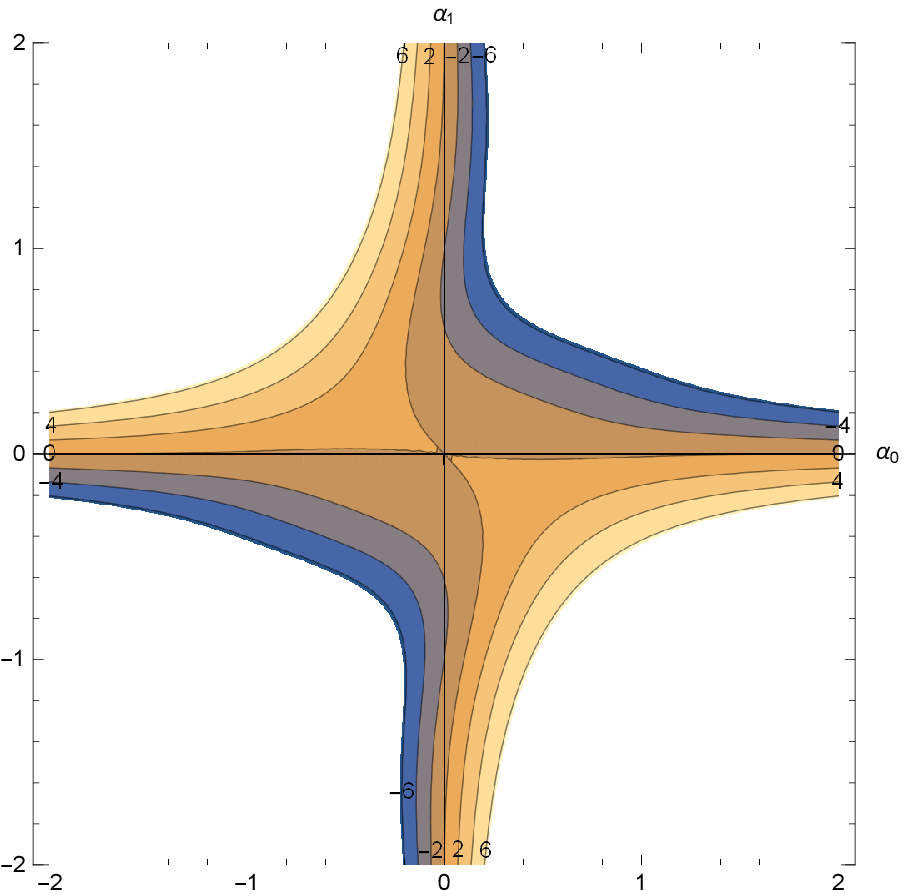}\label{4f}}
 \hfill
  \subfigure[$r_{0}=0.4$, $r_{1}=0$]{\includegraphics[width=5cm]{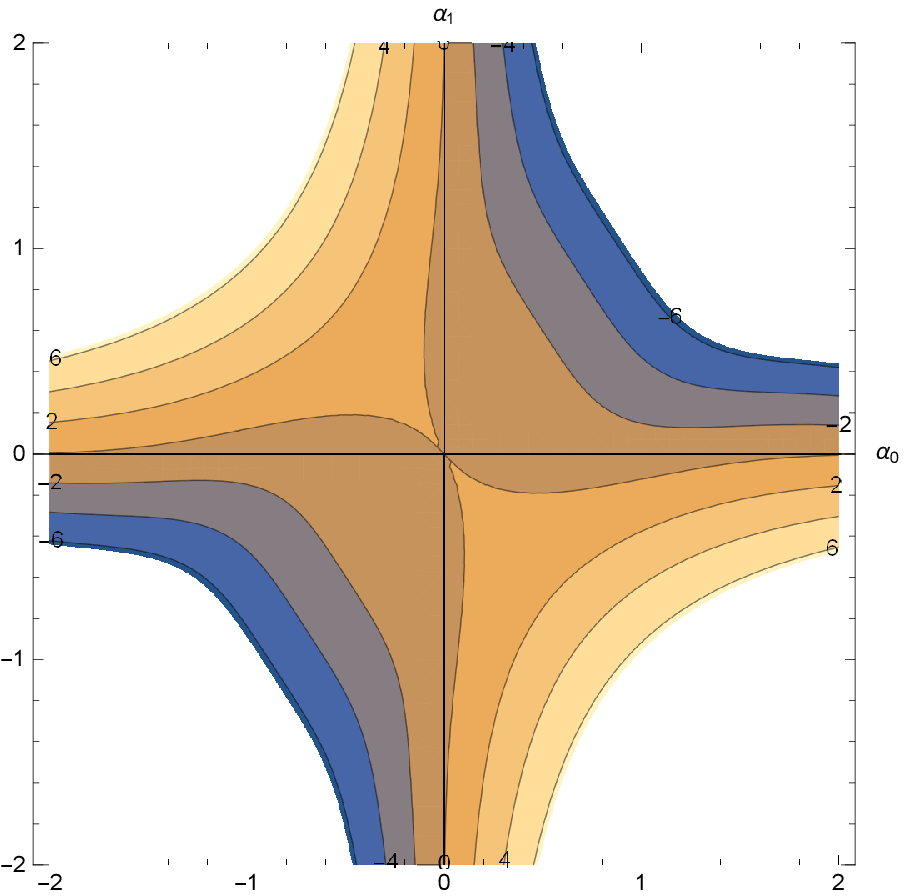}\label{4g}}
 \hfill
  \subfigure[$r_{0}=0.8$, $r_{1}=0$]{\includegraphics[width=5cm]{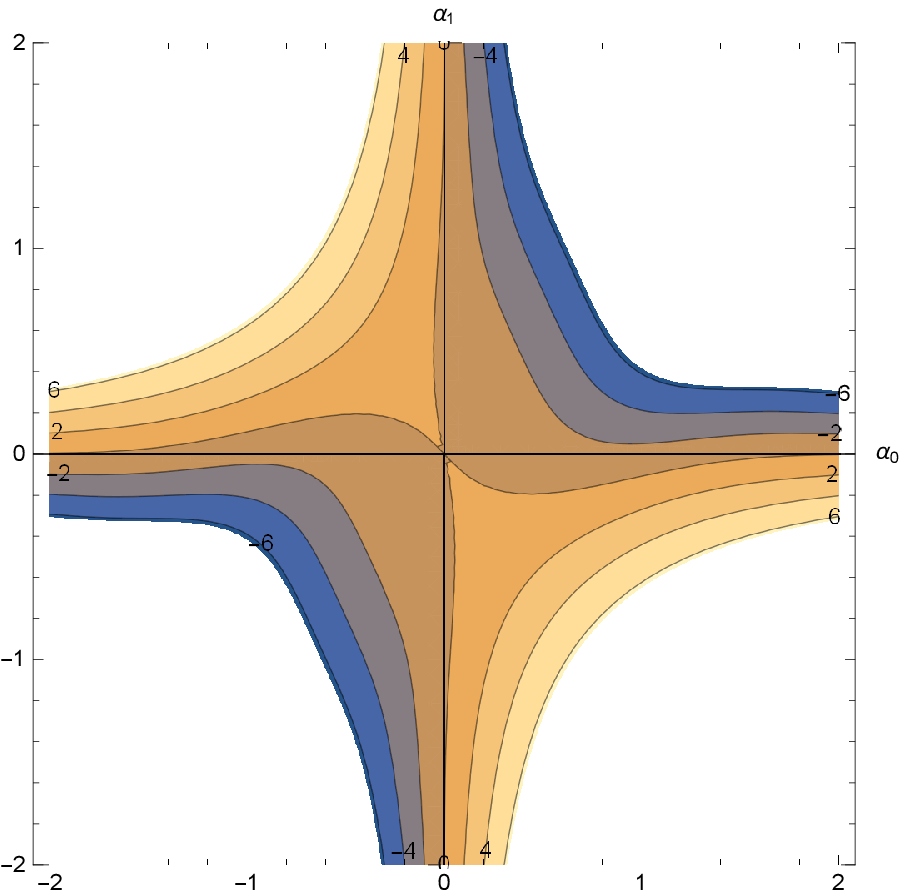}\label{4h}}
 \hfill
  \subfigure[$r_{0}=1.2$, $r_{1}=0$]{\includegraphics[width=5cm]{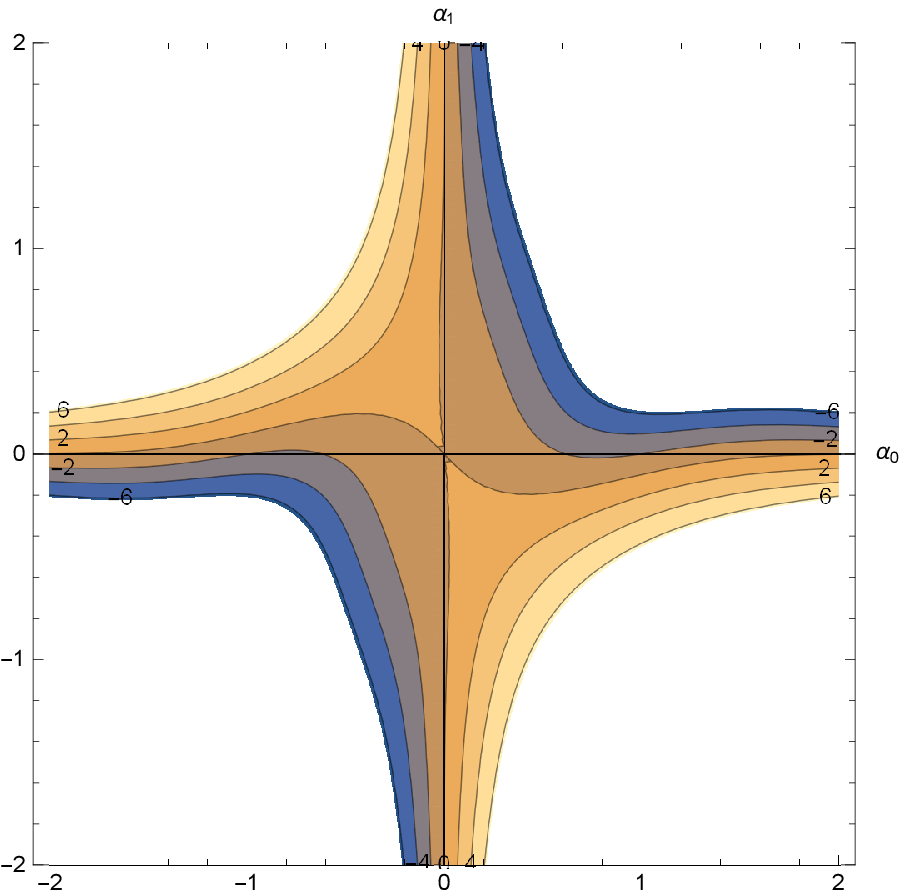}\label{4i}}
  \caption{\small{(Color online.) Contour plots of the GP of $|\psi(0)\rangle_{unbal}$ as functions of $\alpha_{0}$ and $\alpha_{1}$, for $\theta=\pi/4$ and for different values of the squeezing parameters.}}\label{ContourgpunbalESCS}
 \end{figure}

In Fig. \ref{ContourgpunbalESCS}, the calculated GPs of the unbalanced ESCS are plotted as functions of $\alpha_{0}$ and $\alpha_{1}$, for different values of the squeezing parameter. From the figure, it can be seen that, by increasing the squeezing parameters, the plots are compressed in a hyperbolic manner.

\section{Generalization to higher dimensions}
\par
In this section, we proceed to investigate the effect of subsystems dimension on the GP. Let's assume the initial state is a $d$-dimensional two-mode balanced ESCS, i.e.:
\begin{equation}\label{balESCS-d}
|\psi^{(d)}(0)\rangle_{bal}=\frac{1}{\sqrt{M_{bal}^{(d)}}}\sum\limits_{i=0}^{d-1}|\alpha_{i},\xi_{i}\rangle|\alpha_{i},\xi_{i}\rangle,
\end{equation}
where, $M_{bal}^{(d)}=\sum\limits_{i,j=0}^{d-1} p_{ij}^{2}$ is the normalization factor, with $p_{ij}=\langle\alpha_{i},\xi_{i}|\alpha_{j},\xi_{j}\rangle$. Note that the superscript $(d)$ in $|\psi^{(d)}(0)\rangle_{bal}$ and $M_{bal}^{(d)}$ refers to dimension of subsystems. It is clear that for $d=2$, the state $|\psi^{(d)}(0)\rangle$ is the balanced ESCS, defined in Eq. (\ref{balESCS}). 
\par
Applying the unitary evolution operator of Eq. (\ref{unitaryevolution}), to the state $|\psi^{(d)}(0)\rangle$ we get:
\begin{equation}\label{evolved-balESCS-d}
|\psi^{(d)}(\theta,\varphi)\rangle_{bal}=\frac{1}{\sqrt{M_{bal}^{(d)}}}\sum\limits_{i=0}^{d-1}|e^{-i\frac{\varphi}{2}}\alpha_{i}(\cos\frac{\theta}{2}-\sin\frac{\theta}{2}),r_{i}\rangle|e^{i\frac{\varphi}{2}}\alpha_{i}(\cos\frac{\theta}{2}+\sin\frac{\theta}{2}),r_{i}\rangle.
\end{equation}
For a fixed value of $\theta$ and for the cyclic evolution of $\varphi$ from $0$ to $2\pi$, the total acquired phase is equal to zero. The overlap of the initial and final states is a positive and real-valued number, so, the total phase vanishes.

\par
To calculate the GP, we need to calculate the expectation value of $\hat{U}^{\dag}(\theta,\varphi)\partial_{\varphi}\hat{U}(\theta,\varphi)$ for the initial state $|\psi^{(d)}(0)\rangle_{bal}$:
\begin{equation}\label{x-balESCS-d}
_{bal}\langle\psi^{(d)}(0)|\hat{U}^{\dag}(\theta,\varphi)\partial_{\varphi}\hat{U}(\theta,\varphi)|\psi^{(d)}(0)\rangle_{bal}=i \sin\theta \langle\psi^{(d)}(0)|\hat{J_{x}}|\psi^{(d)}(0)\rangle.
\end{equation}
where, $_{bal}\langle\psi^{(d)}(0)|\hat{J_{z}}|\psi^{(d)}(0)\rangle_{bal}=0$ and $_{bal}\langle\psi^{(d)}(0)|\hat{J_{x}}|\psi^{(d)}(0)\rangle_{bal}$ reads as:
\begin{equation}
_{bal}\langle\psi^{(d)}(0)|\hat{J_{x}}|\psi^{(d)}(0)\rangle_{bal}=\frac{1}{M_{bal}^{(d)}}\sum\limits_{i,j=0}^{d-1} \mu_{i}\mu_{j}p_{ij}^{2}\eta_{i}\eta_{j}.
\end{equation}

Putting Eq.(\ref{x-balESCS-d}) in Eq.(\ref{gp}) and by changing $\varphi$ from $0$ to $2\pi$, for a fixed value of $\theta$ the GP is obtained as:
\begin{equation}\label{gp-balESCS-d}
\Phi_{G}(|\psi^{(d)}(0)\rangle_{bal})=-\frac{2 \pi \sin\theta}{M_{bal}^{(d)}}(\sum\limits_{i,j=0}^{d-1} p_{ij}^{2}\eta_{i}\eta_{j}).
\end{equation}
which depends on $\theta$, $d$, and the occupation number of the individual modes of the initial state. Note that, we have assumed all the parameters are real, which yields $\eta=\alpha e^{r}$. For $r=0$, the GP reduces to the GP of the balanced ECSs.
\par

\begin{figure}
  \centering
  \includegraphics[width=12cm]{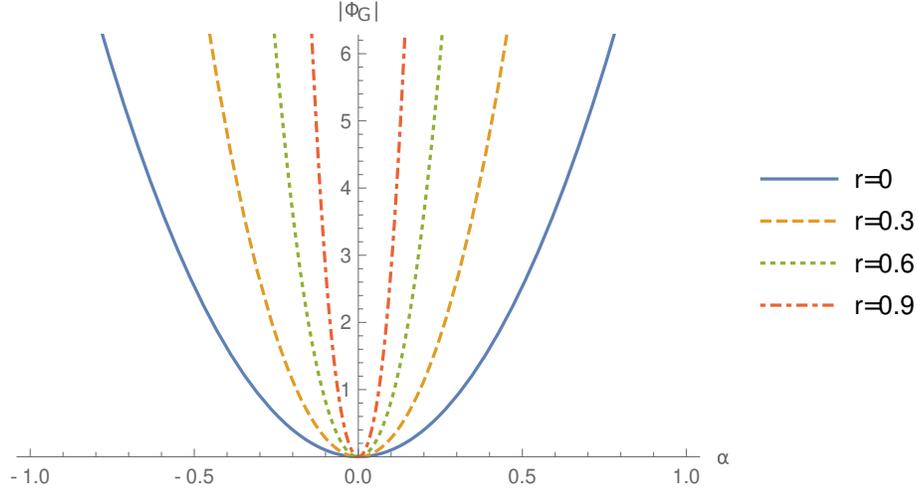}\\
 \caption {\small{The GP of two-mode balanced ESCS as a function of $\alpha$ for $d=2$, $\theta=\pi/4$, and for different values of $r$.}}\label{GP-ECSS-r}
\end{figure}

In Fig. \ref{GP-ECSS-r}, effect of $r$ on the GP is shown for the special case of $\alpha_{i}=(i+1)\alpha$. In this figure, the modulus of the GP is depicted as a function of $\alpha$ for $d=2$, $\theta=\pi/4$, and for different values of $r$. It is clear that we have a symmetry line at $\alpha=0$, i.e., GP is an even function of $\alpha$ and increases by increasing the absolute value of $\alpha$. Also, the GP increases by increasing the value of $r$, for a specific value of $\alpha$. In other words, squeezing intensifies the variations of GP versus $\alpha$.
\par
\begin{figure}
  \centering
  \includegraphics[width=12cm]{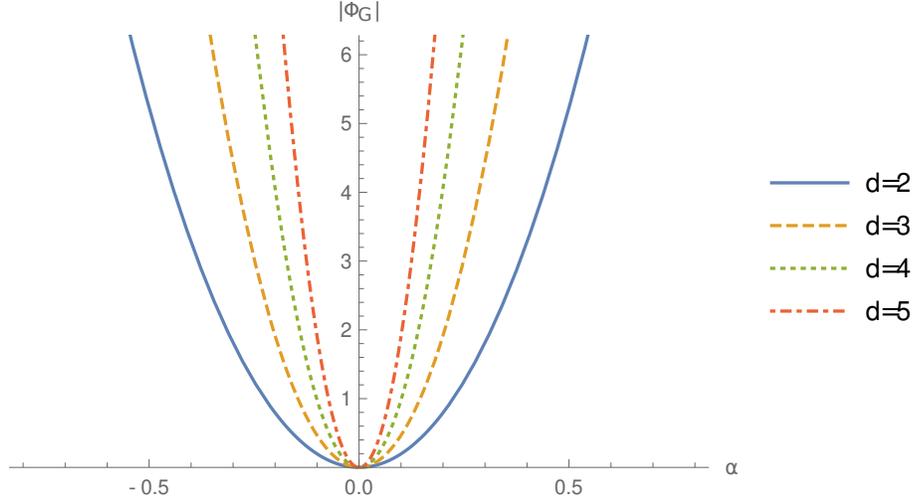}\\
 \caption {\small{The GP of two-mode balanced ESCS as a function of $\alpha$ for $\theta=\pi/4$, $r=0.2$, and for different values of $d$.}}\label{GP-ECSS-d}
\end{figure}

In Fig. \ref{GP-ECSS-d}, the modulus of the calculated GPs are plotted as functions of $\alpha$ for $\theta=\pi/4$, $r=0.2$, and for different values of $d$. This figure helps to understand the effect of dimension on the GP. Here, the diagrams also belong to the special case of $\alpha_{i}=(i+1)\alpha$ and $r_{i}=(i+1)r$. It can be seen that the GP is symmetric with respect to $\alpha=0$, independent of the dimension. It can be seen that the GPs of the states with higher dimensions, increase faster by increasing $\alpha$.

\par
Now, we consider a $d$-dimensional two-mode unbalanced ESCS as
\begin{equation}\label{unbalESCS-d}
|\psi^{(d)}(0)\rangle_{unbal}=\frac{1}{\sqrt{M_{unbal}^{(d)}}}\sum\limits_{i=0}^{d-1} |\alpha_{i},\xi_{i}\rangle|\alpha_{i+1},\xi_{i+1}\rangle,
\end{equation}
with $\alpha_{d}=\alpha_{0}$ and $\xi_{d}=\xi_{0}$. $M_{unbal}^{(d)}=\sum\limits_{i,j=0}^{d-1} p_{ij}p_{(i+1)(j+1)}$ is the normalization factor, with $p_{ij}=\langle\alpha_{i},\xi_{i}|\alpha_{j},\xi_{j}\rangle$. For $d=2$, the state $|\psi^{(d)}(0)\rangle_{unbal}$ reduces to the unbalanced ESCS defined in Eq. (\ref{unbalESCS}). 
\par
Applying the unitary operator of Eq. (\ref{unitaryevolution}), we have: 
\begin{equation}\label{evolved-unbalESCS-d}
|\psi^{(d)}(\theta,\varphi)\rangle_{unbal}=\frac{1}{\sqrt{M_{unbal}^{(d)}}}\sum\limits_{i=0}^{d-1}|(e^{-i\frac{\varphi}{2}}(\alpha_{i}\cos\frac{\theta}{2}-\alpha_{i+1}\sin\frac{\theta}{2})), r_{i}\rangle|(e^{i\frac{\varphi}{2}}(\alpha_{i+1}\cos\frac{\theta}{2}+\alpha_{i}\sin\frac{\theta}{2})),r_{i+1}\rangle.
\end{equation}
For fixed values of $\theta$ and for cyclic evolution of $\varphi$ from $0$ to $2\pi$, total phase vanishes and we have:
\begin{align}\label{gp-unbalESCS-d}
\Phi_{G}(|\psi^{(d)}(0)\rangle_{unbal})=&\frac{ \pi \cos\theta}{M_{unbal}^{(d)}}\{\sum\limits_{i,j=0}^{d-1} p_{ij} p_{(i+1)(j+1)}(\eta_{i}\eta_{j}-\eta_{i+1}\eta_{j+1})\}\\ \nonumber
&-\frac{ \pi \sin\theta}{M_{unbal}^{(d)}}\{\sum\limits_{i,j=0}^{d-1} p_{ij} p_{(i+1)(j+1)}(\eta_{i}\eta_{j+1}-\eta_{j}\eta_{i+1})\}.
\end{align}
\par
Behavior of the obtained GPs of unbalanced ESCS is similar to that of Fig. \ref{GP-ECSS-r} and Fig. \ref{GP-ECSS-d}, for the corresponding special cases. The effect of parameters $r$ and $d$ on the unbalanced GP, for the special case of $\alpha_{i}=(i+1)\alpha$ and $r_{i}=(i+1)r$, is exactly the same as the GP of the balanced ESCS. 

\section{Generation of two-mode ESCSs}
In this section, we suggest a theoretical scheme for generating the state introduced in Eq. (\ref{balESCS}). To generate the two-mode ESCSs, we need a superposition of two squeezed-coherent states like $|\alpha_{0},\xi_{0}\rangle+|\alpha_{1},\xi_{1}\rangle$, up to a normalization factor. A theoretical approach to generate a superposition of distinguishable squeezed-coherent states is devised by involving self-Kerr phase modulation and phase-sensitive parametric amplification \cite{paris1999generation}. In addition, it has been shown that the whole setup is robust against decoherence. With a similar approach, we suggest a simple method to produce the ESCS.
\par
\begin{figure}
  \centering
  \includegraphics[width=10 cm]{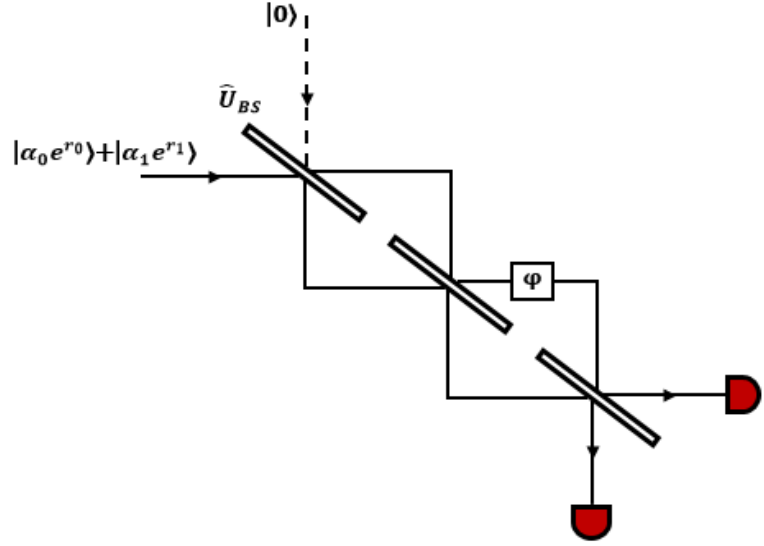}\\
 \caption {\small{Schematic setup to generate the two-mode balanced ESCS.}}\label{generation-ESCS}
\end{figure}

A $50:50$ beam splitter can be defined as a unitary transformation:
\begin{equation}
\hat{U}_{BS}=e^{-i\frac{\pi}{2}\hat{J_{y}}},
\end{equation}
such that the reflected beam acquires a phase shift of $\pi/2$. As depicted in Fig. \ref{generation-ESCS}, the beam splitter acts on an input state that is a superposition of two squeezed-coherent states in one input port and a vacuum state, $|0\rangle$, in the other input port. Therefore, the input state is
\begin{equation}
|\text{input}\rangle=(|\alpha_{0},\xi_{0}\rangle_{1}+|\alpha_{1},\xi_{1}\rangle_{1})\otimes|0\rangle_{2}.
\end{equation}

When the input state passes through the beam splitter, we have
\begin{equation}
\hat{U}_{BS}|\text{input}\rangle=|\frac{\alpha_{0}}{\sqrt{2}}e^{r_{0}}\rangle_{1}|\frac{\alpha_{0}}{\sqrt{2}}e^{r_{0}}\rangle_{2}+|\frac{\alpha_{1}}{\sqrt{2}}e^{r_{1}}\rangle_{1}|\frac{\alpha_{1}}{\sqrt{2}}e^{r_{1}}\rangle_{2},
\end{equation}
redefining $\alpha_{0}\equiv\frac{\alpha_{0}}{\sqrt{2}}$ and $\alpha_{1}\equiv\frac{\alpha_{1}}{\sqrt{2}}$, we can rewrite it as $|\alpha_{0}e^{r_{0}}\rangle_{1}|\alpha_{0}e^{r_{0}}\rangle_{2}+|\alpha_{1}e^{r_{1}}\rangle_{1}|\alpha_{1}e^{r_{1}}\rangle_{2}$, which is a two-mode balanced ESCS. 
We assume the second and the third beam splitters in Fig. \ref{generation-ESCS} perform a rotation around the $y$-axis by the angle $\pi/2$, i.e. $e^{-i\frac{\pi}{2}J_{y}}$, and the phase-shifters perform a rotation around the $x$-axis by the angle $\varphi$, i.e. $e^{-i\varphi J_{x}}$. 
Then, by using $e^{\alpha\hat{A}}f(\hat{B})e^{-\alpha\hat{A}}=f(e^{\alpha\hat{A}} \hat{B} e^{-\alpha\hat{A}}) $, the total action of the three components of the depicted setup can be expressed as
\begin{equation}
e^{i\frac{\pi}{2}J_{y}}e^{-i\varphi J_{x}}e^{-i\frac{\pi}{2}J_{y}}=e^{-i\varphi J_{z}}.
\end{equation}
Finally, the output state can be written as:
\begin{equation}
|\text{output}\rangle=e^{-i\varphi J_{z}}(|\alpha_{0}e^{r_{0}}\rangle_{1}|\alpha_{0}e^{r_{0}}\rangle_{2}+|\alpha_{1}e^{r_{1}}\rangle_{1}|\alpha_{1}e^{r_{1}}\rangle_{2}).
\end{equation}

\section{Summary and Conclusion}
\par
In summary, in this paper, we have studied the geometric phase (GP) of the entangled squeezed-coherent states (ESCSs). The GPs of the two-mode ESCSs are calculated when the states undergo a cyclic unitary evolution. It is revealed that the squeezing parameter compresses the GP functions along the coherence axis. Results show that, for a fixed squeezing parameter of the first (second) mode, the GP of the balanced ESCS compresses by increasing the squeezing parameter of the second (first) mode, in an elliptical manner, along the axis of the coherence parameter of the second (first) mode. For the unbalanced ESCS, calculated GPs are  compressed in a hyperbolic manner, by increasing the squeezing parameters. Also, we have studied the effect of subsystems dimensions on the GP by generalizing it to higher dimensions. We conclude that by putting $\alpha_{i}=(i+1)\alpha$ and $r_{i}=(i+1)r$, the behavior of the GPs for balanced and unbalanced states are the same, i.e., as the dimension increases, the GP increases sharply by increasing $|\alpha|$. Finally, we have suggested a practical scheme to produce the two-mode balanced ESCSs, using an interferometry setup.


\begin{thebibliography}{99}
\bibitem{berry1984quantal}
M. V. Berry, "Quantal phase factors accompanying adiabatic changes," \textit{Proceedings of the Royal Society of London. A. Mathematical and Physical Sciences}, Vol.392, No.1802, pp.45-57, 1984.
\bibitem{wilczek1984appearance}
F. Wilczek and A. Zee, "Appearance of gauge structure in simple dynamical systems," \textit{Physical Review Letters}, Vol.52, No.24, p.2111, 1984.
\bibitem{aharonov1987phase}
Y. Aharonov and J. Anandan, "Phase change during a cyclic quantum evolution," \textit{Physical Review Letters}, Vol.58, No.16, p.1593, 1987.
\bibitem{samuel1988general}
J. Samuel and R. Bhandari, "General setting for Berry's phase," \textit{Physical Review Letters}, Vol.60, No.23, p.2339, 1988.
\bibitem{mukunda1993quantum}
N. Mukunda and R. Simon, "Quantum kinematic approach to the geometric phase. i. general formalism," \textit{Annals of Physics}, Vol.228, No.2, pp.205-268, 1993.
\bibitem{uhlmann1986parallel}
A. Uhlmann, "Parallel transport and quantum holonomy along density operators," \textit{Reports on Mathematical Physics}, Vol.24, No.2, pp.229-240, 1986.
\bibitem{sjoqvist2000geometric}
E. Sj\"{o}qvist, "Geometric phase for entangled spin pairs," \textit{Physical Review A}, Vol.62, No.2,p.022109, 2000.
\bibitem{carollo2003geometric}
A. Carollo, I. Fuentes-Guridi, M. F. Santos, and V. Vedral, "Geometric phase in open systems," \textit{Physical review letters}, Vol.90, No.16, p.160402, 2003.
\bibitem{whitney2003berry}
R. S. Whitney and Y. Gefen, "Berry phase in a Nonisolated system," \textit{Physical Review Letters}, Vol.90, No.19, p.190402, 2003.
\bibitem{tong2004kinematic}
D. Tong, E. Sj\"{o}qvist, L. C. Kwek, and C. H. Oh, "Kinematic approach to the mixed state geometric phase in nonunitary evolution," \textit{Physical Review Letters}, Vol.93, No.8, p.080405, 2004.
\bibitem{klauder1985coherent}
J. R. Klauder and B. S. Skagerstam, "Coherent states: applications in physics and mathematical physics," \textit{World scientific}, 1985.
\bibitem{zhang1990coherent}
W. M. Zhang, R. Gilmore, et al., "Coherent states: theory and some applications," \textit{Reviews of Modern Physics}, Vol.62, No.4, p.867, 1990.
\bibitem{kuratsuji1985effective}
H. Kuratsuji and S. Iida, "Effective action for adiabatic process: Dynamical meaning of Berry and Simon’s phase," \textit{Progress of theoretical physics}, vol.74, no.3, pp.439–445, 1985.
\bibitem{kuratsuji1988geometric}
H. Kuratsuji, "Geometric canonical phase factors and path integrals," \textit{Physical review letters}, vol.61, no.15, p.1687, 1988.
\bibitem{littlejohn1988cyclic}
R. G. Littlejohn, "Cyclic evolution in quantum mechanics and the phases of Bohr-Sommerfeld and Maslov," \textit{Physical review letters}, vol.61, no.19, p.2159, 1988.
\bibitem{mendas1997pancharatnam}
I. Mendas, "Pancharatnam phase for ordinary and generalized squeezed states," \textit{Physical Review A}, vol.55, no.2, p.1514, 1997.
\bibitem{chaturvedi1987berry}
S. Chaturvedi, M. Sriram, and V. Srinivasan, "Berry's phase for coherent states," \textit{Journal of Physics A: Mathematical and General}, Vol.20, No.16, p.L1071, 1987.
\bibitem{pati1995geometric}
A. K. Pati, "Geometric aspects of noncyclic quantum evolutions," \textit{Physical Review A}, Vol.52, No.4, p.2576, 1995.
\bibitem{sjoqvist1997noncyclic}
E. Sj\"{o}qvist, M. Hedstr\"{o}m, “Noncyclic geometric phase, coherent states, and the timedependent variational principle: application to coupled electron-nuclear dynamics,” \textit{Physical Review A}, vol.56, no.5, p.3417, 1997.
\bibitem{yang2011geometric}
D. B. Yang, Y. Chen, F. L. Zhang, and J. L. Chen, "Geometric phases for nonlinear coherent and squeezed states," Journal of Physics B: Atomic, \textit{Molecular and Optical Physics}, Vol.44, No.7, p.075502, 2011.
\bibitem{zanardi1999holonomic}
P. Zanardi and M. Rasetti, "Holonomic quantum computation," \textit{Physics Letters A}, Vol.264, No.2-3, pp.94-99, 1999.
\bibitem{jones2000geometric}
J. A. Jones, V. Vedral, A. Ekert, and G. Castagnoli, "Geometric quantum computation using nuclear magnetic resonance," \textit{Nature}, Vol.403, No.6772, pp.869-871, 2000.
\bibitem{zhu2002implementation}
S. L. Zhu and Z. Wang, "Implementation of universal quantum gates based on nonadiabatic geometric phases," \textit{Physical Review Letters}, Vol.89, No.9, p.097902, 2002.
\bibitem{vedral2003geometric}
V. Vedral, "Geometric phases and topological quantum computation," \textit{International Journal of Quantum Information}, Vol.1, No.01, pp.1-23, 2003.
\bibitem{rowell2018mathematics}
E. Rowell and Z. Wang, "Mathematics of topological quantum computing," \textit{Bulletin of the American Mathematical Society}, Vol.55, No.2, pp.183-238, 2018.
\bibitem{morpurgo1998ensemble}
A. Morpurgo, J. Heida, T. Klapwijk, B. Van Wees, and G. Borghs, "Ensemble-average spectrum of Aharonov-Bohm conductance oscillations: evidence for spin-orbit-induced Berry's phase," \textit{Physical Review Letters}, Vol.80, No.5, p.1050, 1998.
\bibitem{niu1999adiabatic}
Q. Niu, X. Wang, L. Kleinman, W. M. Liu, D. Nicholson, and G. Stocks, "Adiabatic dynamics of local spin moments in itinerant magnets," \textit{Physical Review Letters}, Vol.83, No.1, p.207, 1999.
\bibitem{tiwari1992geometric}
S. Tiwari, "Geometric phase in optics: Quantal or classical?," \textit{Journal of Modern Optics}, Vol.39, No.5, pp.1097-1105, 1992.
\bibitem{galvez2002applications}
E. J. Galvez, "Applications of geometric phase in optics," \textit{Recent Research Developments in Optics}, Vol.2, pp.165-182, 2002.
\bibitem{sjoqvist2015geometric}
E. Sj\"{o}qvist, "Geometric phases in quantum information," \textit{International Journal of Quantum Chemistry}, Vol.115, No.19, pp.1311-1326, 2015.
\bibitem{tong2003geometric}
D. Tong, L. Kwek, and C. Oh, "Geometric phase for entangled states of two spin-1/2 particles in rotating magnetic field," \textit{Journal of Physics A: Mathematical and General},
Vol.36, No.4, p.1149, 2003.
\bibitem{tong2003relation}
D. Tong, E. Sj\"{o}qvist, L. Kwek, C. Oh, and M. Ericsson, "Relation between geometric phases of entangled bipartite systems and their subsystems," \textit{Physical Review A}, Vol.68, No.2, p.022106, 2003.
\bibitem{bertlmann2004berry}
R. A. Bertlmann, K. Durstberger, Y. Hasegawa, and B. C. Hiesmayr, "Berry phase in entangled systems: A proposed experiment with single neutrons," \textit{Physical Review A}, Vol.69, No.3, p.032112, 2004.
\bibitem{najarbashi2017quantum}
G. Najarbashi and B. Seifi, "Quantum phase transition in the Dzyaloshinskii-Moriya interaction with inhomogeneous magnetic field: Geometric approach," \textit{Quantum Information Processing}, Vol.16, No.2, pp.1-16, 2017.
\bibitem{gerry2005introductory}
C. Gerry, P. Knight, and P. L. Knight, "Introductory quantum optics," \textit{Cambridge university press}, 2005.
\bibitem{beals2016special}
R. Beals and R. Wong, "Special functions and orthogonal polynomials," \textit{Cambridge University Press}, vol.153, 2016.
\bibitem{Almas2022geometric}
S. Mohammadi Almas, G. Najarbashi, and A. Tavana, "Geometric phase for two-mode entangled coherent states," \textit{Int J Theor Phys}, vol.61, no.192, 2022.
\bibitem{paris1999generation}
M. G. Paris, "Generation of mesoscopic quantum superpositions through Kerr-stimulated degenerate downconversion," \textit{Journal of Optics B: Quantum and Semiclassical Optics}, vol.1, no.6, p.662, 1999.
\end{thebibliography}
\end{document}